\documentclass{aastex63}
\usepackage{amsmath}
\usepackage{enumitem}
\usepackage[T1]{fontenc}
\usepackage{gensymb}
\usepackage{multirow}

\newcommand\chisq{$\chi^2$}
\newcommand\dchisq{$\Delta\chi^2$}
\newcommand\Dc{\mathcal{D}}
\newcommand\lsun{L$_\sun$}
\newcommand\mearth{M$_\earth$}
\newcommand\msun{M$_\sun$}
\newcommand\No{\mathcal{N}}
\newcommand\prot{P_{\rm rot}}
\newcommand\ptra{p_{\rm tra}}

\newcommand\rearth{R$_\earth$}
\newcommand\rsun{R$_\sun$}
\newcommand\searth{S$_\earth$}
\newcommand\teff{T_{\rm eff}}

\newcommand\Un{\mathcal{U}}

\accepted{April 27, 2023}
\submitjournal{AJ}

\shorttitle{The Occurrence Rate of Terrestrial Planets Orbiting Mid-to-late M Dwarfs}
\shortauthors{Ment \& Charbonneau}

\graphicspath{{./}{figures/}}

\begin{document}

\title{The Occurrence Rate of Terrestrial Planets Orbiting Nearby Mid-to-late M Dwarfs from TESS Sectors 1-42}

\correspondingauthor{Kristo Ment}
\email{kristo.ment@cfa.harvard.edu}

\author[0000-0001-5847-9147]{Kristo Ment}
\affiliation{Center for Astrophysics \textbar~Harvard \& Smithsonian, 60 Garden Street, Cambridge, MA 02138, USA}

\author[0000-0002-9003-484X]{David Charbonneau}
\affil{Center for Astrophysics \textbar~Harvard \& Smithsonian, 60 Garden Street, Cambridge, MA 02138, USA}

\begin{abstract}

We present an analysis of a volume-complete sample of 363 mid-to-late M dwarfs within 15 pc of the Sun with masses between 0.1-0.3 \msun~observed by TESS within Sectors 1 to 42. The median stellar mass of the sample is 0.17 \msun. We search the TESS light curves for transiting planets with orbital periods below 7 days and recover all 6 known planets within the sample as well as a likely planet candidate orbiting LHS 475. Each of these planets is consistent with a terrestrial composition, with planet radii between 0.91-1.31 \rearth. We characterize the transit detection sensitivity for each star as a function of planet radius, insolation, and orbital period. We obtain a cumulative occurrence rate of $0.61^{+0.24}_{-0.19}$ terrestrial planets per star with radii above 0.5 \rearth~and orbital periods between 0.4-7 days. We find that for comparable insolations, planets larger than 1.5 \rearth~(sub-Neptunes) are significantly less abundant around mid-to-late M dwarfs compared to earlier-type stars, while the occurrence rate of terrestrial planets is comparable to that of more massive M dwarfs. We estimate that overall, terrestrials outnumber sub-Neptunes around mid-to-late M dwarfs by 14 to 1, in contrast to GK dwarfs where they are roughly equinumerous. We place a $1\sigma$ upper limit of 0.07 planets larger than 1.5 \rearth~per star, within the orbital period range of 0.5-7 days. We find evidence for a downturn in occurrence rates for planet radii below 0.9\rearth, suggesting that Earth-sized and larger terrestrials may be more common around mid-to-late M dwarfs.

\end{abstract}


\section{Introduction} \label{sec:intro}

In the field of terrestrial planet studies, M dwarfs are the most lucrative planet hosts. They are more common than Sun-like stars, they appear to have a higher incidence rate of orbiting planets \citep{Dressing2013,Bonfils2013,Mulders2015}, and their planets are easier to detect and study due to their larger size compared to the host star. Apart from white dwarfs, they are nature's best targets for transmission and emission spectroscopy \citep{Morley2017} to detect and characterize exoplanetary atmospheres. Furthermore, large quantities of precise photometric data of M dwarfs have recently become available following the launch of the Transiting Exoplanet Survey Satellite (TESS) in 2018. Previous terrestrial planet population studies have largely relied on data yielded from the Kepler and K2 missions \citep[e.g.][]{Dressing2015} that were designed to study Sun-like stars. Kepler observed few nearby mid-to-late M dwarfs that are sufficiently bright for small transiting planets to be detected. The motivation for the current study is to take advantage of TESS photometry to constrain the presence of close-in terrestrial planets among mid-to-late M dwarfs, the least massive stars studied in this manner to date, and to place these findings in a broader context by comparing them to previous results for early M as well as G and K dwarfs.

\citet{Dressing2015} estimate a cumulative planet occurrence rate of $2.5 \pm 0.2$ planets per M dwarf with planet radii 1-4 \rearth~and orbital periods below 200 days, based on a sample of early-to-mid M dwarfs with a median stellar radius of 0.47 \rsun. They report an occurrence rate of $0.69$ close-in planets per star, defined by radii of 0.5-4 \rearth~and orbital periods between 0.5-10 days. Intriguingly, the incidence rate of planets orbiting GK dwarfs as a function of radius is bimodal, with a gap in the distribution at 1.5-2 \rearth~\citep{Fulton2017}. These empirical findings supported previous studies that had found two distinct populations of small planets: those with radii $r<1.5R_\earth$ consistent with a rocky bulk composition (so-called "super-Earths"), and planets with $r>1.5R_\earth$ that possess large volatile envelopes \citep["mini-Neptunes",][]{Weiss2014,Rogers2015}. A similar bimodality in the planet radius distribution has also been identified in a distribution of planet-hosting mid-K to mid-M dwarfs, with the center of the "radius valley" shifting towards smaller sizes with decreasing stellar mass \citep{Cloutier2020}. More recently, \citet{Luque2022} found evidence of three planet populations orbiting M dwarfs (rocky, water-rich, and gas-rich) separated by mean density as opposed to planet radius.

The radius distribution of the occurrence rate has direct implications for theories of planet formation. In particular, several hypotheses have emerged as possible explanations to the existence of the radius valley, including photoevaporation \citep{Lopez2013,Owen2013,Owen2017,Lopez2018}, core-powered mass loss \citep{Ginzburg2018}, and formation in a gas-poor environment \citep{Lee2014,Lopez2018}. \citet{Luque2022} propose that water-rich exoplanets initially form beyond the snow line and later migrate inwards. While mini-Neptunes and super-Earths appear to be nearly equal in abundance around Sun-like stars \citep{Fulton2017}, terrestrial planets increase in relative prominence around M dwarfs \citep{Cloutier2020}.

The relative abundance of sub-Earths (planets with sizes less than $r<1R_\earth$) compared to super-Earths has been poorly constrained. The vast majority of terrestrial exoplanets discovered thus far orbit M dwarfs and are larger than Earth; meanwhile, every terrestrial planet in the Solar System is Earth-sized or smaller. Consequently, the total mass of known terrestrial planets in many M dwarf planetary systems is comparable to or exceeds the total mass of the inner Solar System. This is surprising since M dwarfs are much less massive than the Sun and therefore possessed less massive protoplanetary disks in their early evolution \citep[e.g.][]{Pascucci2016}. Owing to their relative large size compared to planets, mid-to-late M dwarfs are bound to be the first type of star where detection sensitivity allows the underlying sub-Earth incidence to be studied in earnest. In addition to calculating cumulative occurrence rates, this work seeks to constrain the relative abundance of sub-Earths compared to Earth-sized and larger terrestrials.

\section{Stellar sample and data}\label{sec:stars}

\begin{figure}
    \includegraphics[width=\textwidth]{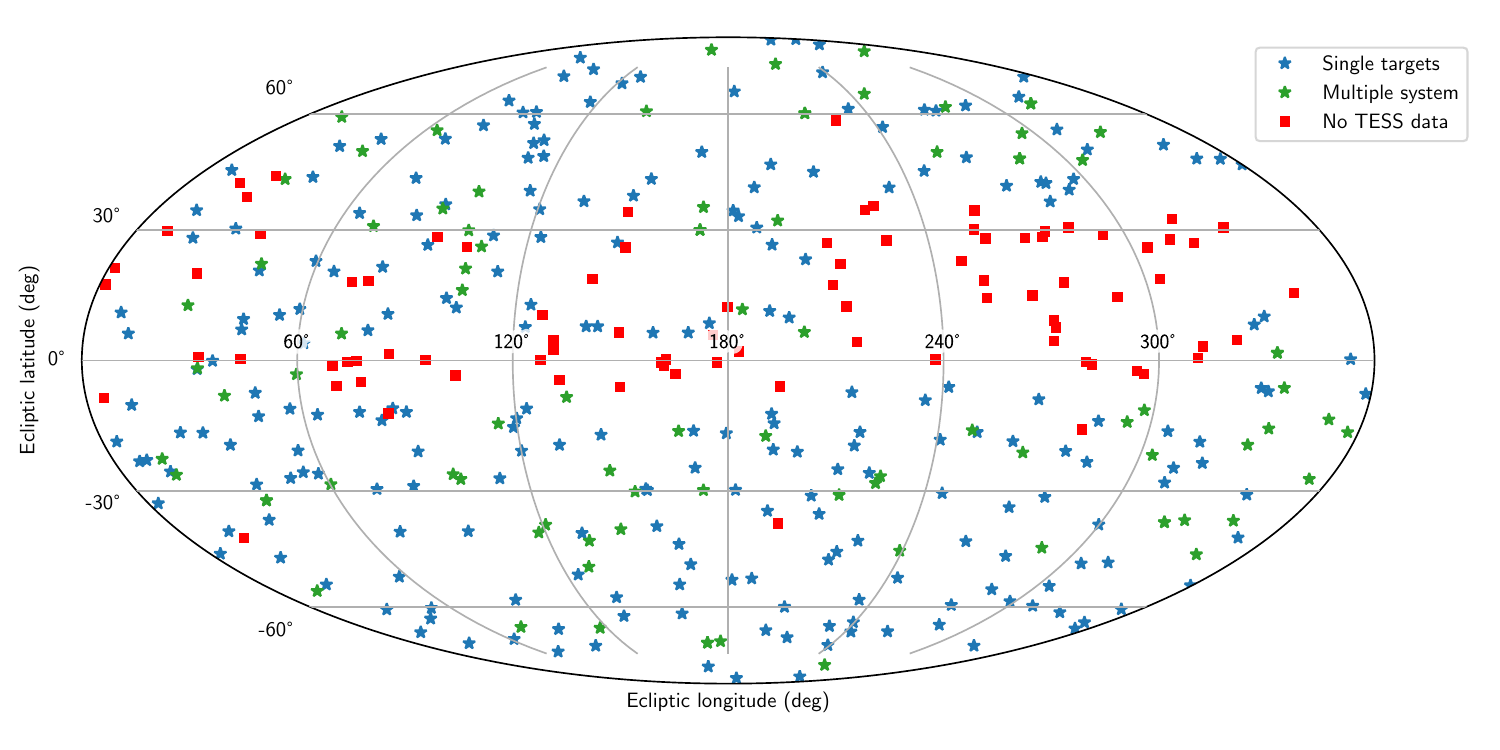}
    \caption{Spatial sky distribution in ecliptic coordinates of the volume-complete 15 pc M dwarf sample, excluding close blended binaries where the primary star is not an M dwarf. Star symbols mark the locations of the stars studied in this article, including single M dwarfs (blue) and M dwarfs in multiple systems (green). Red squares denote stars that would have otherwise been included in this study, but did not have any available data in observation sectors 1-42 due to gaps in spatial coverage.}
    \label{fig:coord}
\end{figure}

We begin by considering the volume-complete, all-sky sample of 512 mid-to-late M dwarfs within 15 pc of the Sun listed by \citet{Winters2021}. This sample covers the stellar mass range $0.1 \leq M / M_\sun \leq 0.3$, corresponding roughly to spectral types between M4V and M7V. We retain the single M dwarfs as well as multiples where the mid-to-late M dwarf is either the primary star or widely separated from the primary star (separations greater than 4"), yielding a total of 413 targets. In addition, 60 of our mid-to-late M dwarf primaries have close companions (separations $<$4") that are also mid-to-late M dwarfs. We include those close companions in our sample for a total of 473 stars. We subsequently narrow our search down to the 363 stars that were observed by TESS in observation sectors 1 through 42. All of these targets appear in the TESS Input Catalog (TIC) and many were included in the Candidate Target List \citep[CTL;][]{Stassun2018}. Close binaries generally share the same TIC ID. The spatial distribution of the targets included in this study is presented in Figure \ref{fig:coord}.

We utilize the publicly available two-minute cadence Presearch Data Conditioning \citep[PDCSAP;][]{Smith2012,Stumpe2012,Stumpe2014} light curve data generated by the NASA Ames Science Processing Operations Center (SPOC) pipeline \citep{Jenkins2016}. These data have been calibrated and processed to remove common instrumental systematics as well as corrected for additional flux in the aperture due to crowding from nearby stars. The entire data set spans the time period from July 25, 2018 to September 14, 2021. When downloading the data, we adjust the median flux in each sector to be zero for every star as well as reject impulsive outliers as defined by SPOC (quality flag 512).

\begin{figure}
    \includegraphics[width=\textwidth]{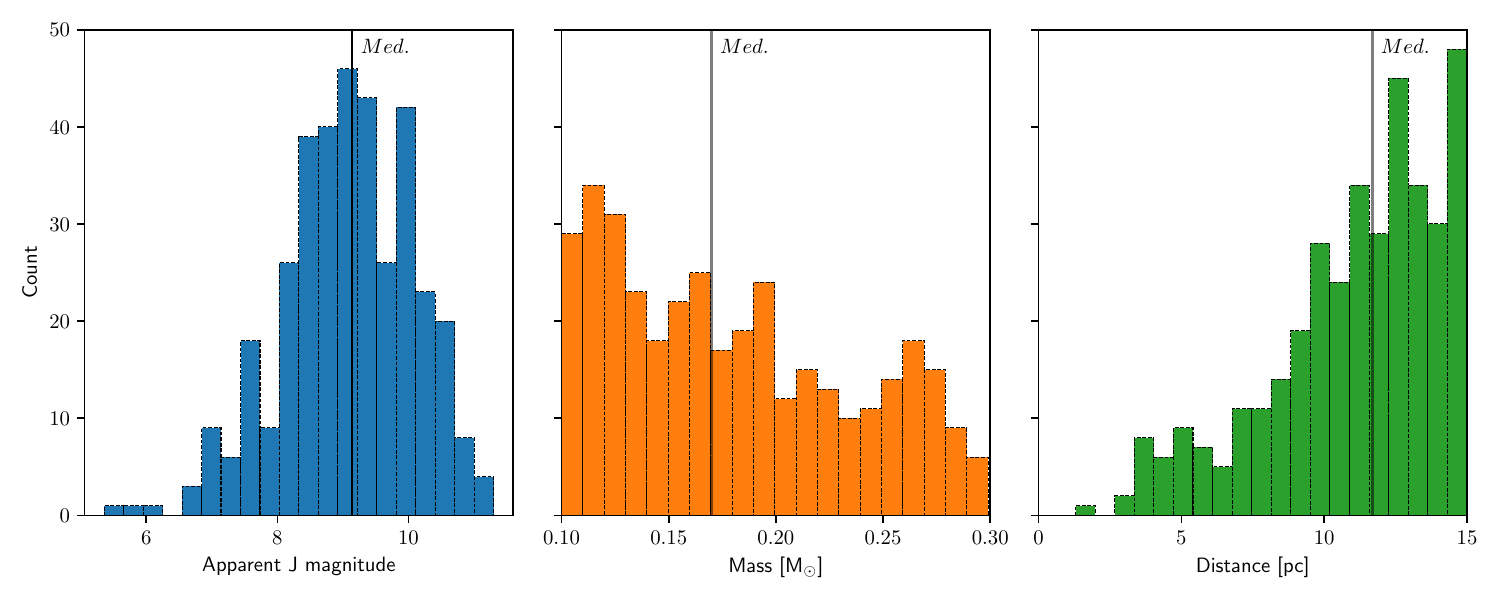}
    \caption{Distribution of apparent J magnitudes, masses, and distances for the stars in our sample. Vertical lines denote median values. For blended systems with multiple M dwarfs, we include each star separately using estimated deblended magnitudes.}
    \label{fig:stardist}
\end{figure}

We adopt stellar masses from \citet{Winters2021}. We then proceed to estimate each star's radius from a mass-radius relation calibrated via optical interferometry \citep{Boyajian2012}. Stellar luminosities are obtained via a bolometric correction: we employ the third-order polynomial fit between $BC_J$ and $V-J$ in \citet[and its erratum]{Mann2015}. The $J$ fluxes, masses, and distances of the stars in our sample are illustrated in Figure \ref{fig:stardist}, and numerical values are given in Table \ref{tbl:stars}. We also assume quadratic limb-darkening coefficients for each star in the TESS photometric system from Table 15 of \citet{Claret2018}. These coefficients are based on the spherical PHOENIX-COND limb darkening model \citep{Husser2013} using the appropriate values for $\teff$ and $\log g$, which we estimate from the stellar mass, radius, and luminosity.

For close binaries with blended photometry, we proceed as follows. The stellar masses in \citet{Winters2021} were deblended before mass estimation calculations so we simply adopt these. Furthermore, Table 3 of \citet{Winters2021} also lists magnitude differences ($\Delta m$) between multiple components, albeit in various different filters. We obtain the corresponding differences in $V$ and $J$ bandpasses using conversions produced by \citet{Riedel2014}. This allows us to find luminosities for both of the companions (in the case of binary stars). The TESS PDCSAP light curves for close binaries are generally blended as well; in these cases, we deblend the fluxes using the estimated magnitude difference in the TESS bandpass ($\Delta T$) and produce a separate light curve for each M dwarf in our sample before proceeding with the analysis. We estimate $\Delta T$ by calculating the TESS magnitudes $T$ for each component from $(I-K)$ colors as described in \citet{Winters2019}. A complete list of stars included in this work is provided in Table \ref{tbl:stars}.

\begin{deluxetable}{cccccccccccc}
    \tabletypesize{\footnotesize}
	\tablewidth{0pt}
	\tablecaption{Properties of target stars\label{tbl:stars}}
	\tablehead{
		TIC & Sectors & 2MASS ID & Comp.\tablenotemark{a} & Sep.\tablenotemark{b} & Dist. & $V$ & $J$ & Mass & Radius & Luminosity & $\prot$\\
		 & & & & ('') & (pc) & (mag) & (mag) & (\msun) & (\rsun) & (\lsun) & (days)
	}
	\startdata
	 101955023 & 9,10,36 & 10145184$-$4709244 & \nodata & \nodata & 12.62 & 13.49 & 9.245 & 0.192 & 0.219 & 0.00443 & \nodata\\
	 102286828 & 6,33 & 05565722$+$1144333 & \nodata & \nodata & 14.25 & 15.72 & 10.369 & 0.125 & 0.171 & 0.00185 & \nodata\\
	 103245015 & 14,41 & 20000565$+$3057300 & \nodata & \nodata & 11.65 & 16.49 & 10.674 & 0.102 & 0.156 & 0.00091 & \nodata\\
	 111898820 & 39 & 17464063$-$3214045 & \nodata & \nodata & 12.06 & 16.15 & 10.350 & 0.116 & 0.165 & 0.00132 & 0.229\\
	 114308371 & 7,34 & 07365666$-$3024160 & \nodata & \nodata & 13.90 & 13.67 & 9.359 & 0.196 & 0.222 & 0.00481 & \nodata\\
	\enddata
	\tablenotetext{a}{Component identifier, if the star is in a multiple system.}
	\tablenotetext{b}{Separation between components in a multiple system.}
	\tablecomments{For members of multiple systems with blended photometry, the values for $V$ and $J$ magnitudes, masses, radii, and luminosities are estimated separately for each component, as described in Section \ref{sec:stars}. Table \ref{tbl:stars} is published in its entirety in machine-readable format.}
\end{deluxetable}

Furthermore, we split the data for each target M dwarf such that every light curve encompasses at most 2 consecutive TESS sectors (at most 58 days). This is done out of computational cost considerations for the Box-Least Squares analysis described in Section \ref{sec:trdetection} that scales non-linearly with the span of the time baseline. As a result, most stars in our study have multiple light curves that undergo analysis independently of each other, and we combine these results later. Finally, we note that our sample includes five M dwarfs that are known to host transiting planets: GJ 1132 \citep[TIC 101955023;][]{BertaThompson2015,Bonfils2018}, LHS 1140 \citep[TIC 92226327;][]{Dittmann2017,Ment2019}, LHS 3844 \citep[TIC 410153553;][]{Vanderspek2019}, LTT 1445A \citep[TIC 98796344;][]{Winters2019,Winters2022}, and TOI 540 \citep[TIC 200322593;][]{Ment2021}.

\section{Methods}\label{sec:methods}

\subsection{Outlier rejection and detrending}\label{sec:detrend}

Before proceeding with our analysis, we clean the 2-minute cadence light curves as follows. We begin by visually inspecting all light curves for any evidence of anomalies that can likely be attributed to residual instrumental systematics or data quality issues and are not astrophysical in nature. We then manually exclude the corresponding chunks of data from further analysis. This typically covers rapid flux fluctuations at the beginning or the end of a spacecraft orbit that indicate potential calibration issues. In some cases, we also manually redact data related to powerful stellar flares that are not appropriately removed by the automated flare-detection algorithm described below. All in all, this manual removal process affects 26\% of the stars in our sample (for the remaining 74\% of stars, no data were removed), and typically no more than 5\% of an individual light curve is excluded in this manner. We note that while redacting data can adversely affect detection sensitivity, it does not bias the final results as transits are unlikely to occur preferentially during periods of poor data.

Subsequently, we analyze each light curve with a Lomb-Scargle (LS) periodogram to seek evidence of periodic activity likely resulting from stellar rotation. Before any such period is accepted, it must pass a series of quality tests, including a requirement that the power of the highest LS peak be at least 3 times that of the next most significant peak. The latter requirement is waived for targets that are binary stars that show two clearly distinguishable peaks in the periodogram; in those cases, both of the periods are accepted. The resulting "rotation" periods are included in Table \ref{tbl:stars}\footnote{While many of these periods likely correspond to actual rotation periods, they have not been rigorously vetted to be classified as such. These periods are only adopted as "rotation" periods for detrending purposes.}.

For single targets where a rotation period is detected, we proceed by modeling the flux baseline with a Gaussian Process (GP) using the Python/Theano implementation of \textit{celerite2} \citep{ForemanMackey2017,ForemanMackey2018}. We adopt a covariance kernel that is the sum of a quasiperiodic rotation term (a mixture of two damped harmonic oscillators, SHO) and an exponentially decaying term:
\begin{equation}\label{eq:kernel}
    k(\tau) = k_{\rm Rotation}(\tau; Q_0, \Delta Q, \prot, \mathcal{A}, \lambda) + k_{\rm Mat\acute{e}rn}(\tau; \sigma, \rho)
\end{equation}
The quasiperiodic first half of $k(\tau)$ is equivalent to the Rotation kernel implemented in the \textit{celerite2} package. It consists of a mixture of two stochastically driven damped harmonic oscillators; we describe the functional form of this kernel in \citet{Ment2021}. The exponentially decaying term is an approximation of the Mat\'ern-3/2 kernel, for a small value of $\epsilon$ (here, $\epsilon = 0.01$):
\begin{equation}
    k_{\rm Mat\acute{e}rn}(\tau; \sigma, \rho) = \frac{\sigma^2}{2} \left[ \left(1 + \frac{1}{\epsilon}\right) e^{-(1-\epsilon)\sqrt{3}\tau/\rho} + \left(1 - \frac{1}{\epsilon}\right) e^{-(1+\epsilon)\sqrt{3}\tau/\rho} \right]
\end{equation}
\begin{equation}
    \lim_{\epsilon \to 0} k_{\rm Mat\acute{e}rn}(\tau; \sigma, \rho) = \sigma^2 \left(1 + \frac{\sqrt{3}\tau}{\rho}\right) e^{-\sqrt{3}\tau/\rho}
\end{equation}
The full baseline model allows for the inclusion of a constant term $\mu$ to be added to the flux. It also includes a constant white noise parameter $\sigma_{\rm WGN}$ that is added to all individual flux uncertainties in quadrature. All parameters are further constrained by prior probability distributions, including a tight constraint on the rotation period $\prot$. We provide a complete list of the parameters that we optimize in Table \ref{tbl:params}.

\begin{deluxetable}{ccccc}
    \tabletypesize{\footnotesize}
	\tablewidth{0pt}
	\tablecaption{Model parameters for detrending\label{tbl:params}}
	\tablehead{
		Parameter & Explanation & Prior & Bounds & Units
	}
	\startdata
	$\mu$ & Flux baseline & $\No(0, 10)$ & - & mmag\\
	$\ln \sigma_{\rm WGN}^2$ & Excess white noise & $\No(\ln \sigma_{\rm RMS}^2, 5)$ & - & mmag$^2$\\
	$Q_0$ & Quality parameter & $\No(15\pi - 0.5, 0.03\pi)$ & $(15\pi - 0.5, \infty)$ & -\\
	$\ln \Delta Q$ & Quality parameter & $\No(0, 5)$ & - & -\\
	$\prot$ & Rotation period & $\No(P_{\rm LS}, 0.01P_{\rm LS})$ & - & days\\
	$\mathcal{A}$ & Covariance amplitude & $\Un(0, \sigma_{\rm RMS}^2)$ & $(0, \sigma_{\rm RMS}^2)$ & mmag$^2$\\
	$\lambda$ & Covariance amp. ratio & $\Un(0, 1)$ & $(0, 1)$ & -\\
	$\ln \sigma^2$ & Covariance amplitude & $\Un(0, \sigma_{\rm RMS}^2)$ & $(0, \sigma_{\rm RMS}^2)$ & mmag$^2$\\
	$\ln \rho$ & Covariance decay timescale & $\No(1.648, 0.173)$ & $(1.648, \infty)$ & days\\
	\enddata
	\tablecomments{$\No(\mu, \sigma)$ denotes a normal distribution. $\Un(a, b)$ denotes a uniform distribution. $\sigma_{\rm RMS}^2$ is the standard deviation of the initial light curve. $P_{\rm LS}$ is the accepted rotation period from the LS periodogram.}
\end{deluxetable}

For binary targets where the LS periodogram detects two non-overlapping peaks, we use an additional rotation term in the covariance kernel centered at the other detected rotation period. This term is added to the kernel in Equation \ref{eq:kernel} and is identical in form to the Rotation kernel in Equation \ref{eq:kernel}.

Nearly two thirds of our targets do not show evidence for substantial quasi-periodic modulation. In these cases, we adopt a sliding median window with a 12-hour width to model the baseline. This is intended to capture any long-term variations in flux while being sufficiently coarse not to remove any detectable planetary transits which have a typical duration of 1 hour or less.

After optimizing the baseline fit, we perform an automated flare detection analysis to redact any data with rapid variations in flux due to stellar flaring. A flare is defined as a sequence of at least four consecutive data points that are brighter than 2$\sigma$ or more compared to the fitted baseline. Here, $\sigma$ denotes the quoted individual uncertainties for each data point. Once a flare is identified, all positive deviants relative to the fitted baseline surrounding the detected flare are redacted until the observed flux returns to normal. In addition, we generate a binned version of the light curve (in bins of 0.1 days) and redact any part of it where the RMS of the flux residuals is at least 170\% of the median value for all bins. The 170\% threshold was chosen to exclude time periods with unusually high flux variations.

The above cleaning procedure (including the baseline fitting) is repeated until no additional data is redacted in two successive iterations, or at most four times.

\subsection{Detection and removal of planetary transits}\label{sec:trdetection}

We conduct a modified Box-Least Squares \citep[BLS;][]{Kovacs2002,Burke2006} search to detect any transiting planets. The BLS algorithm is set up to look for planets with orbital periods between 0.1 and 7 days. We limit the analysis to periods below 7 days to decrease the computational cost of the transit injection and recovery phase (see Section \ref{sec:inject}). The geometric transit probability at 7 days for a median star in our sample is only 2.5\%. Thus, we would only expect a small number of detected planets with orbital periods above 7 days in a sample of 363 stars even if the underlying occurrence rate was relatively high. The number of such detections would ultimately be too low to place statistically meaningful constraints on the underlying planet incidence rate. In addition, limiting the orbital period to below 7 days guarantees that a single TESS sector covers at least 3 transits for any tested model. The orbital search frequencies are evenly log-spaced such that the resulting orbital phase shift for each data point between two successive search frequencies is guaranteed to be less than 4.4 minutes. In addition, the BLS runs on time-binned phase-folded data. The duration of each time bin is set to 10.08 minutes or less such that the total number of bins remains at least 100. The resulting BLS spectra are represented in tested frequency vs. \dchisq, the difference in \chisq~between the best-fitting box model and a constant flux model at the corresponding orbital frequency.

We analyze a total of 752 light curves, produced by splitting the time series data into chunks containing at most 2 consecutive TESS sectors as described in Section \ref{sec:stars}. We identify three candidate periods for each light curve, corresponding to the three highest peaks in the BLS spectrum. We calculate the effective signal-to-noise ratio (SNR) of each identified BLS peak, defined as the square root of the BLS power:
\begin{equation}\label{eq:snr}
    \rm SNR \equiv \sqrt{\Delta\chi^2}
\end{equation}
We note that for a constant Gaussian noise level $\sigma$, $\sqrt{\Delta\chi^2} = (\delta / \sigma) \sqrt{N_t}$, where $\delta$ is the transit depth and $N_t$ is the number of in-transit data points. We then refit each transit signal with a limb-darkened transit model at the appropriate orbital phase and period, for a total of 2256 models. The parameters of the fitted model are shown in Table \ref{tbl:params_pl}. We assume a circular orbit due to the fact that planetary orbits with periods studied here (less than 7 days) are likely to be tidally circularized, and slight deviations from zero eccentricity will not substantially alter the shape of the transit. The resulting transit models are then saved and subtracted from the data. We make a reasonable assumption that there are no more than 3 detectable true planetary transit signals in each light curve; we thereby also surmise that the resulting data contain no further transits. This assumption is supported by the fact that no more than 2 transits (typically 0) ultimately pass the transit vetting process described below for any light curve. We also note that given the short duration (compared to the orbital period) and fixed shape of a planetary transit, subtracting the three most promising modelled transits from the data does not substantially alter the noise properties of the residual light curve.

\begin{deluxetable}{ccccc}
    \tabletypesize{\footnotesize}
	\tablewidth{0pt}
	\tablecaption{Model parameters for limb-darkened transits\label{tbl:params_pl}}
	\tablehead{
		Parameter & Explanation & Prior & Units
	}
	\startdata
	$\mu$ & Flux baseline & $\No(1, 0.1)$ & -\\
	$P$ & Orbital period & $\No(P_{\rm BLS}, 0.007)$ & days\\
	$t_0$ & Transit mid-point & $\No(t_{0, \rm BLS}, 0.007)$ & JD\\
	$r/R$ & Planet-to-star radius ratio & $\Un(0, 0.1)$ & -\\
	$b$ & Impact parameter & $\Un(0, 1 + r/R)$ & -\\
	\enddata
	\tablecomments{$\No(\mu, \sigma)$ denotes a normal distribution. $\Un(a, b)$ denotes a uniform distribution.}
\end{deluxetable}

\subsection{Residual noise characterization}\label{sec:noise}

\begin{figure}
    \includegraphics[width=\textwidth]{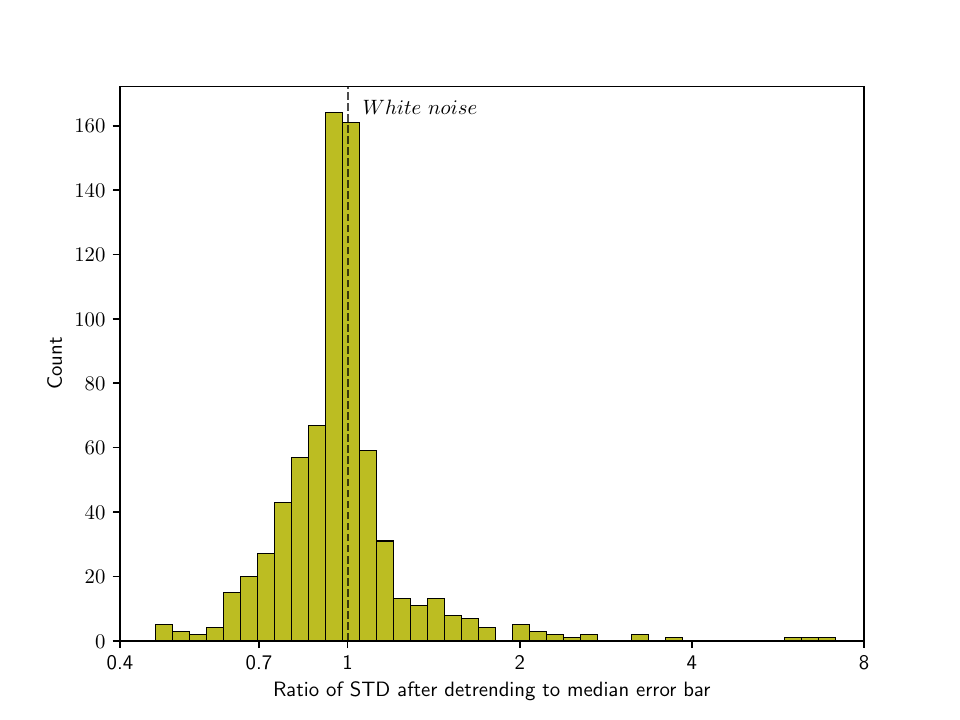}
    \caption{Distribution of ratios of standard deviations of flux (STD) to median error bar after detrending, for each analyzed light curve. For pure white noise, this value is expected to be 1 if the error bars reflect the random uncertainty in the data. Values substantially different from unity may lead to errors in estimating the signal-to-noise ratio of any planetary transit, assuming the data have been successfully detrended to resemble white noise. The median ratio of the distribution is 0.97, suggesting that there is no systematic error in the estimated noise in the sample as a whole, but quoted uncertainties may still be under- or overestimated within individual light curves.}
    \label{fig:residstd}
\end{figure}

After the removal of the flux baseline, outliers, stellar flares, and potential planetary transits, we note that the vast majority of the residual noise in the data appears to be white, with no visible time-correlated structure. However, the root mean square (RMS) of the detrended light curves can be substantially different from the median value of individual uncertainties assigned by the SPOC pipeline, as illustrated in Figure \ref{fig:residstd}. Thus, the quoted individual uncertainties may in some cases yield inaccurate estimates for the amount of white noise present in the data. This can subsequently lead to a misinterpretation of the true significance of any potential planetary transit event. To address this issue, we derive an individual noise scaling factor for each light curve by calculating a BLS peak power distribution of the detrended light curve.

We phase-fold every residual light curve obtained in Section \ref{sec:trdetection} using 2000 equally log-spaced periods between 0.1 and 7 days. We then evaluate the SNR (Eq. \ref{eq:snr}) of a box-shaped transit model as a function of orbital phase, for a large number of orbital phases at every trial period. For every orbital phase, we draw a value for the impact parameter $b$ from a uniform distribution between 0 and 1, which sets the duration of the box-shaped transit at the given period. We then allow the transit depth to vary to maximize the value of \dchisq. We discard all orbital phases with SNR values corresponding to anti-transits (negative transit depths) to be consistent with our transit vetting algorithm in Section \ref{sec:vetting}. We fit the remaining set of SNR values at each orbital period with a half-normal distribution (since SNR values are positive) and record the scale $\sigma_P$ of the distribution for every trial period $P$. For perfectly white noise with each individual measurement uncertainty equal to the standard deviation of the noise, we expect $\sigma_P = 1$ regardless of period $P$. We note that for most light curves, we indeed find at best a weak correlation between $\sigma_P$ and $P$, supporting the assumption that the residual data is mostly comprised of white noise. Consequently, we adopt the median $\sigma_P$ of each light curve as its noise scaling coefficient $\sigma'$, and define the noise-scaled SNR of any planetary transit model as follows:
\begin{equation}\label{eq:noise}
    \rm{SNR}' = \frac{1}{\sigma'} \sqrt{\Delta\chi^2}
\end{equation}
The different values of $\sigma'$ range from 0.44 to 6.51 depending on the light curve, suggesting that the amount of residual noise in the data, if purely white, is either overestimated (for $\sigma'<1$) or underestimated (for $\sigma'>1$) by the quoted flux uncertainties.

\subsection{Transit vetting}\label{sec:vetting}

The BLS algorithm is always guaranteed to find a best-fit box-shaped model, and therefore additional vetting needs to be carried out to eliminate false positives. To this end, we subject all transit models (TM) from Section \ref{sec:trdetection} to a series of quality tests designed to flag low-significance events and/or common sources of false positives. In particular, any transit candidate event (TCE) from BLS must pass the following series of tests:
\begin{enumerate}[label=(\alph*)]
    \item The noise-scaled SNR' must be at least 6.39 if the light curve contains one TESS sector's worth of data, and at least 6.55 for two TESS sectors. We arrived at these values by randomly selecting 5 light curves from each sample (1 or 2 sectors worth of data), replacing the data by sampling from a normal distribution with a fixed variance ($\sigma' = 1$) while adopting the quoted time stamps, and calculating the SNR' for the highest BLS peak. This simulation was repeated 10,000 times per light curve. The aforementioned limits correspond to equal-to-or-greater-than SNR' values that appear with an average frequency of $1/N$ per light curve, where $N$ is the total number of stars studied. Thus, if all data were composed of white noise, this test would yield 1 false positive for the entire sample.
    \item After subtracting a limb-darkened version of the TM from the data, we evaluate the noise-scaling coefficient $\sigma_P$ at the fixed orbital period $P$ of the TM following the process outlined in Section \ref{sec:noise}. Replacing $\sigma'$ with $\sigma_P$ in Equation \ref{eq:noise}, the new SNR' must still be above the SNR' cutoff of test (a). This ensures that there are no other significant transit-like signals with the same orbital period but a different orbital phase.
    \item We calculate $\Delta\chi_{\rm even}^2$ and $\Delta\chi_{\rm odd}^2$ by including only even- and odd-numbered planetary orbits, respectively. The two values must be consistent with each other within a factor of 2.
    \item Data from any single transit event must contribute less than 70\% of the total \dchisq~of the TM.
    \item The \chisq~of the TM must be lower than the \chisq~of any sinusoidal model with period $P/i$, where $P$ is the trial orbital period and $i \in \{1, 2, 3\}$.
    \item The fitted full duration of the transit (using a box model) cannot exceed the maximum allowed transit duration for a circular orbit at the given period and stellar parameters by more than 40\%.
    \item If the given target has light curves from multiple sets of sectors, there must exist a TM with an orbital period within 1\% of the given period that also satisfies the SNR' requirement of test (a) in each of the light curves. This ensures that the signal is present in all observed sectors.
    \item While keeping the ingress and egress times fixed and refitting the transit depth separately for each transit event, an event in considered \textit{significant} if the fitted depth is at least 50\% of the TM depth. Among all transit events, the fraction of significant events cannot be less than 2/3.
\end{enumerate}

All TMs and vetting results are also visually inspected to ensure that the vetting tests are working as intended. We note that tests (a) and (b) alone eliminate over 98\% of the TMs. Furthermore, all but one TM that passed the first three tests (a), (b), and (c) also passed the remaining 5 tests (d)-(h), increasing our confidence that the TCEs represent real planets rather than false positives. Thus, tests (d)-(h) serve primarily as additional sanity checks. We describe the full set of obtained TCEs in Section \ref{sec:planets}.

\subsection{Transit injection and recovery}\label{sec:inject}

In order to estimate our transit detection sensitivity as a function of planet radius and orbital period, we undertake a planet injection and recovery analysis for each individual light curve. We constrain our simulation to orbital periods between 0.1 and 7 days (for reasons covered in Section \ref{sec:trdetection}) and planet-to-star radius ratios $r/R$ between 0 and 0.1 (corresponding to $r<1.9$ \rearth~for the median star). Should a need arise to evaluate sensitivity for $r/R > 0.1$, we simply adopt the value at the same orbital period and $r/R = 0.1$, which is typically close to 100\% in every tested light curve.

We initially generate a $4 \times 4$ sensitivity grid for every light curve by subdividing the parameter space evenly in $\ln P$ and $r/R$. Within each cell, we generate 50 light curves by injecting transit signals from simulated planets. While the number of injected transits per cell was dictated by computational cost, we examined the consistency of our results by repeating the simulation for multiple light curves using different sets of randomly drawn planets. In addition, we increased the number of simulated planets per cell to 100 for multiple light curves and found no major discrepancies in the resulting sensitivity grid. For each simulated planet, the orbital period $P$ is drawn from a distribution that scales linearly with transit duration, $p(P) \propto P^{-2/3}$. The transit phase $t_0 \sim \Un(0, P)$, the impact parameter $b \sim \Un(0, 1)$, and the radius ratio $r/R$ are drawn from uniform distributions. The orbital eccentricity is fixed to 0 for all injections. We generate each simulated light curve by adding a limb-darkened transit model (with appropriate quadratic limb-darkening coefficients for the target star) to a modified version of the initial light curve. This modified version is equivalent to the original light curve, save for the fact that limb-darkened transit models from the 3 TMs from Section \ref{sec:trdetection} have been subtracted from it. This ensures the absence of any additional transit signals caused by real planets in the data, which would bias the subsequent sensitivity calculation.

For each simulated light curve, we perform a mock transit recovery analysis by first cleaning and detrending the data in accordance with the procedure outlined in Section \ref{sec:detrend}. We then obtain three new TMs from the detrended light curve by following the BLS setup in Section \ref{sec:trdetection}. We consider the simulated transit signal recovered contingent on at least one of the TMs meeting both of the following conditions:
\begin{enumerate}[label=(\alph*)]
    \item The noise-scaled SNR' must be at least 6.39 or 6.55, if the light curve contains one or two sectors of TESS data, respectively. This is equivalent to test (a) of Section \ref{sec:vetting}.
    \item The recovered period $P'$, or a multiple $nP'$ or $P'/n$ of the recovered period, where $n \in \{2, 3, 4, 5, 6, 7\}$, must be within 0.2\% of the true orbital period $P$. The average likelihood of a randomly chosen period to pass this test is about 1\%.
\end{enumerate}

\begin{figure}
    \includegraphics[width=\textwidth]{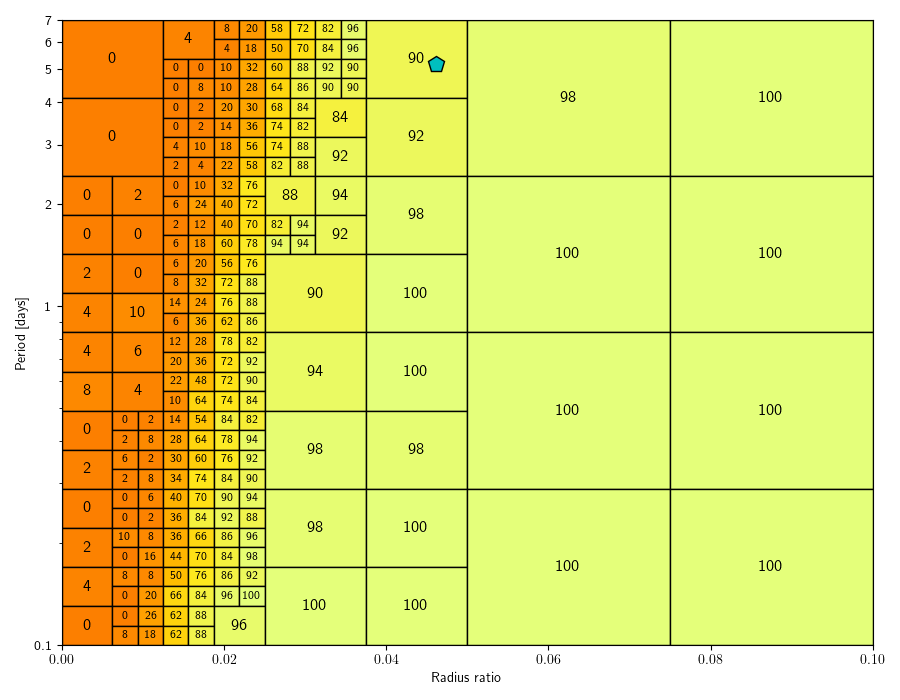}
    \caption{Example detection sensitivity map for a single light curve (TIC11893637). Each cell contains 50 injected planets, with the displayed numeric values corresponding to the percentage of these planets that are successfully recovered. This star has a radius of 0.199 \rsun~and the smallest cells span a planet radius range of 0.06 \rearth. Detection sensitivity generally increases with planet size and orbital frequency. The blue pentagon marker represents an Earth-size planet with surface insolation equal to Earth's.}
    \label{fig:exsens}
\end{figure}

The sensitivity in each grid cell is defined simply as the fraction of recovered planets (out of 50). The grid itself is structured and stored as a quadtree\footnote{A quadtree is a tree data structure in which each internal node has exactly four children.}, with each cell corresponding to a child node. Hence, each cell occupies $(1/4) \times (1/4)$ of the two-dimensional parameter set $\{\ln P\} \times \{r/R\}$. We proceed by splitting any cell into four equal-size child cells when the following the conditions are met:
\begin{enumerate}[label=(\roman*)]
    \item The sensitivity of the given cell differs from that of a neighboring cell by 20\% or more.
    \item Each of the resulting child nodes occupies no less than $(1/32) \times (1/32)$ of the entire parameter space.
\end{enumerate}
This process is repeated until there are no more cells for which the splitting conditions hold true. After each splitting, the 50 injections stored in the parent node are redistributed among the four child nodes, and each child node is then subsequently filled to hold 50 injections by simulating additional transit signals. Therefore, the total number of injected transits in a finished grid is a number between $4 \times 4 \times 50 = 800$ and $32 \times 32 \times 50 = 51,200$. An example of a sensitivity grid for a typical light curve is shown in Figure \ref{fig:exsens}.

\subsection{Cumulative planet occurrence rates}

In reality, the planet occurrence rate $q(P, r)$ varies as a function of both orbital period $P$ and planet radius $r$. Unlike previous major studies on occurrence rates of terrestrial planets from transit photometry \citep{Dressing2015,Fulton2017}, our small sample size of planet detections (described in detail in Section \ref{sec:planets}) limits our ability to place meaningful limits on the exact distribution of $q(P, r)$. Instead, we adopt simple functional forms for $q$ parameterized by a set of parameters $\{\theta\}$ such that the models described the observed distribution reasonably well. We then proceed to generate results under the assumption that the underlying model is valid. For all parameterizations, we assume that $q$ can be separated into a radius-dependent and period-dependent component:
\begin{equation}\label{eq:occrate}
    q(P, r | \{\theta\}) \propto f(P | \{\theta\}) \times g(r | \{\theta\})
\end{equation}
Tested functional forms for $q(P, r | \{\theta\})$ are described in Section \ref{sec:occparam}. Since we divide the parameter space up in $\ln P$ and $r/R$, we define $f(P) = \frac{dN}{d(\ln P)}$ and $g(r) = \frac{dN}{dr}$ where $N$ is the underlying expected number of planets.

For every star $k$, the \textit{success probability} $w$ of observing a planet with an orbital period $P$ and radius $r$ given an underlying occurrence rate $q(P, r)$ is given by the following expression:
\begin{equation}
    w_k(P, r | \{\theta\}) \sim q(P, r | \{\theta\}) \times S_k(P, r) \times \ptra(P, R_k, M_k)
\end{equation}
where $S_k(P, r)$ is the detection sensitivity determined in Section \ref{sec:inject} and $\ptra(P, R, M)$ is the geometrical transit probability given stellar radius $R$ and mass $M$. For stars with multiple light curves, we adopt the highest value of $S_k(P, r)$ from our simulations.

We estimate the best-fit values for the set of parameters $\{\theta\}$ within each bin by sampling $w_k(P, r)$ on a fixed grid of $P$ versus $r$ and maximizing the likelihood of observing the given data set $\Dc$:
\begin{equation}\label{eq:occrate2}
    p(\{\theta\} | \Dc) \propto p(\Dc | \{\theta\}) = \prod_i \prod_j \prod_k w_k(P_i, r_j)^{\delta_k(P_i, r_j)} \times (1 - w_k(P_i, r_j))^{1 - \delta_k(P_i, r_j)}
\end{equation}
where $\delta_k(P_i, r_j) = 1$ if the star $k$ possesses a planet with orbital period $P_i$ and radius $r_j$, and $\delta_k = 0$ otherwise. Since the latter is true for the vast majority of cells, the above equation can be more conveniently expressed as follows:
\begin{equation}
    p(\Dc | \{\theta\}) = \left( \prod_i \prod_j \prod_k 1 - w_k(P_i, r_j)\right) \times \prod_t \frac{w_t(P_t, r_t)}{1 - w_t(P_t, r_t)} = p(\Dc_0 | \{\theta\}) \prod_t \frac{w_t(P_t, r_t)}{1 - w_t(P_t, r_t)}
\end{equation}
where $p(\Dc_0 | \{\theta\})$ is the likelihood of observing 0 planets in the entire sample, and the product to the right is evaluated over the set of accepted TCEs $\{t\}$. The grid cells are made small enough ($(1/64) \times (1/64)$ of the entire parameter space) that we can safely assume no star would contain more than one planet within each grid cell. 

We obtain posterior distributions for each of the model parameters $\theta$ through static nested sampling with the \textit{dynesty} module \citep{Speagle2020,Skilling2004,Skilling2006,Feroz2009}. The uncertainties quoted in Section \ref{sec:results} represent the 15.9th, 50th, and 84.1st percentiles of the posterior PDFs (1$\sigma$ uncertainty), or the 68.2nd percentile of the posterior PDFs (1$\sigma$ upper limit) in the case of small values that are consistent with zero. In addition, we evaluate the total model evidence $Z$ by integrating over all model parameters, allowing us to rank competing models for $q(p, r)$. The cumulative occurrence rate $Q$ over some range of orbital periods $P$ and planet radii $r$ is obtained via a simple sum over the corresponding grid space:
\begin{equation}
    Q(\{\theta\}) = \sum_i \sum_j q(P_i, r_j | \{\theta\})
\end{equation}

\section{Results}\label{sec:results}

\subsection{Detected planets}\label{sec:planets}

From a total of 2256 TMs from Section \ref{sec:trdetection}, only 18 pass all of the vetting tests listed in Section \ref{sec:vetting}. This includes duplicate entries where the same planet is independently detected in multiple TESS sectors or belonging to different components of a multiple star system with blended light curves. After amalgamating the duplicate entries, we obtain 8 unique TCEs: six corresponding to previously known planets (GJ 1132 b, LHS 1140 c, LHS 3844 b, LTT 1445A b and c, TOI 540 b), one corresponding to a known TOI (910.01), and one previously unidentified signal around AP Col (TIC 160329609). We ultimately discard the TCE of AP Col: the detected period is one fourth of the stellar rotation period and in-depth analysis of individual transits reveals systematic inconsistencies in the shape and depth of individual transit events, including an incompatible transit duration for one out of every four transits. Thus, we conclude that it is likely a false positive induced by periodic flux modulation due to stellar rotation. We note that there exist no additional published planets or TOIs in the rest of the sample.

\begin{figure}
    \centering
    \includegraphics[width=0.7\textwidth]{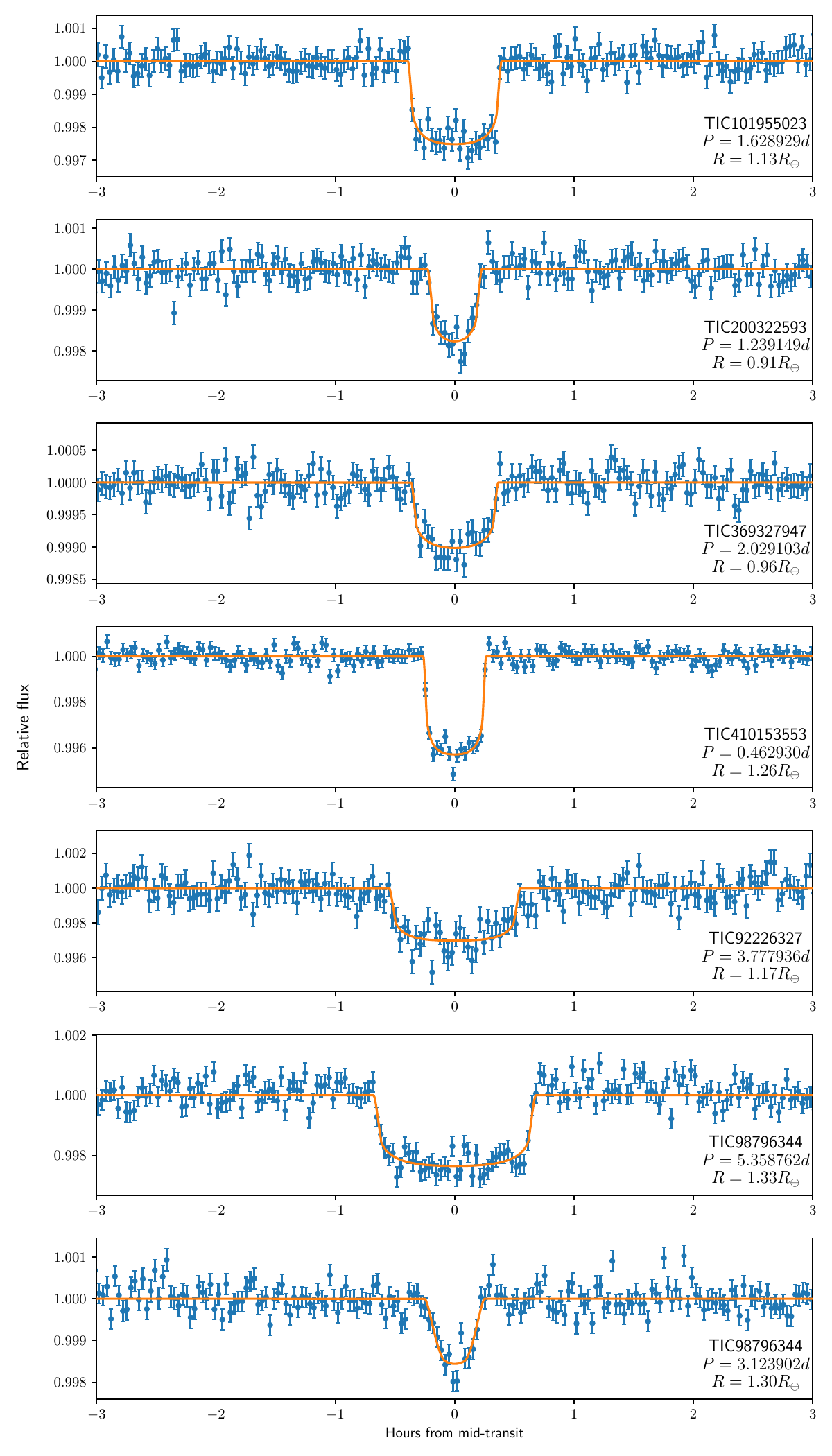}
    \caption{Limb-darkened transit models of the 7 accepted candidate planets: (from top to bottom) GJ 1132 b, TOI 540 b, TOI 910.01, LHS 3844 b, LHS 1140 c, LTT 1445A b, and LTT 1445A c. The orbital eccentricity is set to 0 in each of these models. The individual data points and uncertainties represent the phase-folded light curve binned to 2 minutes per data point.}
    \label{fig:ldtransits}
\end{figure}

We create limb-darkened models for the 7 accepted planets using combined data from all the analyzed sectors. The results are displayed in Figure \ref{fig:ldtransits}. We begin by cleaning and flattening the light curves using either a sliding median or a Gaussian process as described in Section \ref{sec:detrend}. Transit modeling is carried out with \textit{exoplanet} using a Keplerian orbit with orbital eccentricity fixed to 0. We include tight Gaussian priors on the orbital period and phase, corresponding to the highest peak in the BLS spectrum of the combined data. The obtained periods and radii are consistent with previously published values, except for LTT 1445A c where we find a slightly inflated radius of 1.30 \rearth~compared to $1.15^{+0.06}_{-0.05}$ \rearth~by \citet{Winters2022}. Since the \citet{Winters2022} study includes ground-based follow-up data in addition to the entirety of the TESS photometry analyzed here, we adopt the radii of the two planets orbiting LTT 1445A from \citet{Winters2022} for the remainder of the analysis. We provide a list of final parameters for the seven accepted planets in Table \ref{tbl:planets}.

We note that our initial sample of 512 stars contains two additional known planets that are not included as part of the analysis in this work. While the TESS data does contain transits of LHS 1140 b, it has an orbital period of 24.7 days \citep{Dittmann2017,Ment2019}, much longer than our cutoff of 7 days. Therefore, we simply redacted any data associated with these transits within in the light curve of LHS 1140. In addition, GJ 1214 b is a super-Earth with an orbital period of 1.58 days and a radius of 2.7 \rearth~\citep{Charbonneau2009}; however, GJ 1214 is not among the set of 363 stars that had been observed by TESS at the time of this study.

\begin{deluxetable}{lccccc}
    \tabletypesize{\footnotesize}
	\tablewidth{0pt}
	\tablecaption{List of stars with accepted TCEs; planet parameters.\label{tbl:planets}}
	\tablehead{
	    TIC & Name & Sectors & Period & Radius & Max. SNR'\tablenotemark{a}\\
	     & & & (days) & (\rearth) &
	}
	\startdata
	101955023 & GJ 1132 & 9,10,36 & 1.628929 & 1.13 & 21.48\\
	200322593 & TOI 540 & 4,5,6,31,32 & 1.239149 & 0.91 & 12.78\\
	369327947 & LHS 475 & 12,13,27,39 & 2.029103 & 0.96 & 15.71\\
	410153553 & LHS 3844 & 1,27,28 & 0.462930 & 1.26 & 40.47\\
	92226327 & LHS 1140 & 3,30 & 3.777936 & 1.17 & 15.35\\
	98796344 & LTT 1445A & 4,31 & 5.358762 & 1.31\tablenotemark{b} & 19.32\\
	98796344 & LTT 1445A & 4,31 & 3.123902 & 1.15\tablenotemark{b} & 11.69\\
	\enddata
	\tablenotetext{a}{Highest individual SNR' from all the TMs (from different sectors) that were used to identify the accepted planet.}
	\tablenotetext{b}{Adopted from \citet{Winters2022}.}
\end{deluxetable}

\subsection{Survey sensitivity}\label{sec:surveysens}

\begin{figure}
    \includegraphics[width=\textwidth]{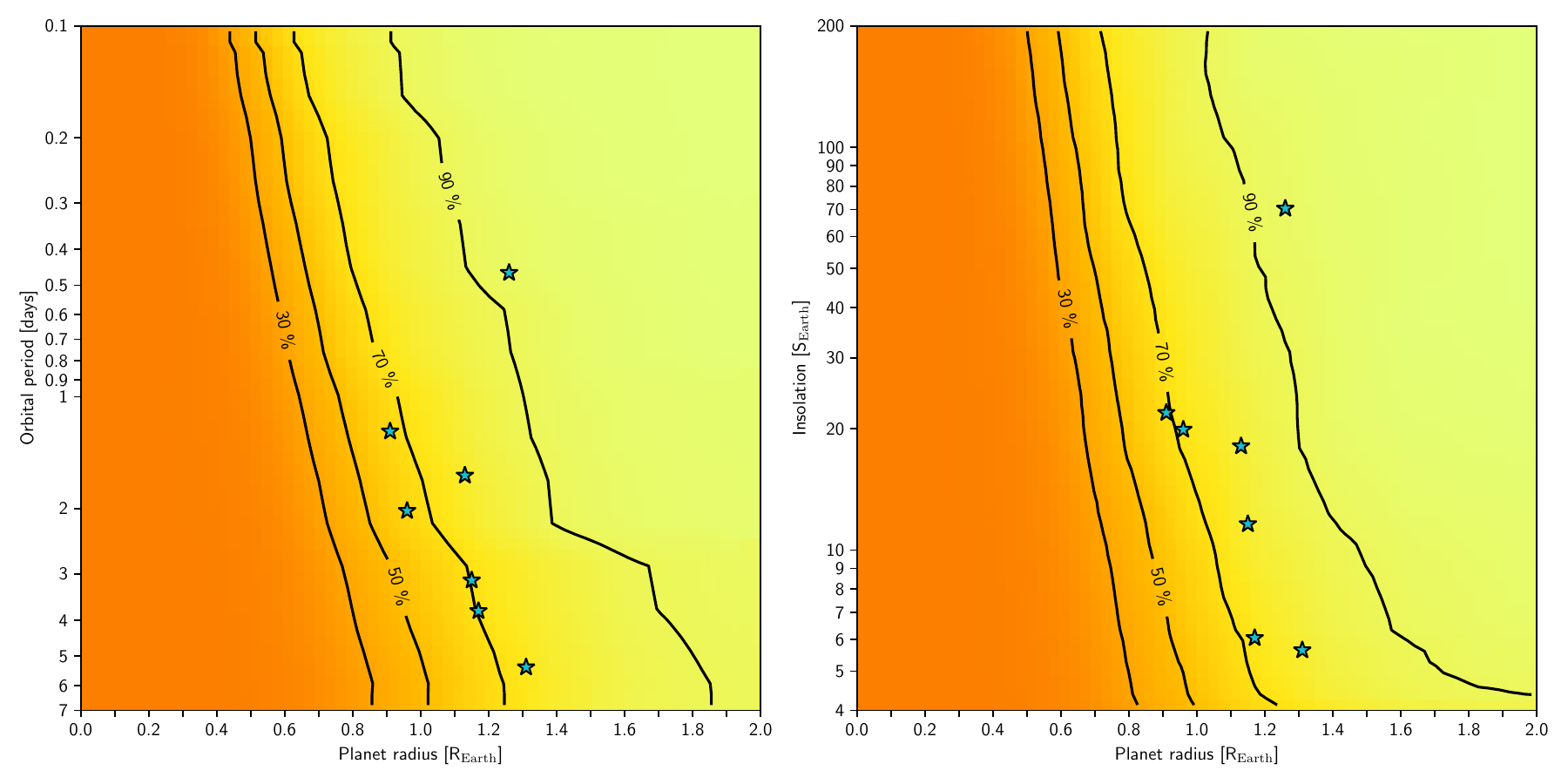}
    \caption{The average transit detection sensitivity as a function of planet radius and either orbital period (left) or stellar insolation (right). Contour lines are drawn to mark the 30\%, 50\%, 70\%, and 90\% sensitivity thresholds. In addition, the locations of the 7 planets are marked by stars.}
    \label{fig:surveysens}
\end{figure}

In addition to the sensitivity maps described in Section \ref{sec:inject}, we produce a survey sensitivity map that is the sum of all the sensitivity grids for individual targets (for stars with multiple sensitivity grids, we use the one with the highest sensitivity) divided by the number of targets. Therefore, it represents the average survey sensitivity for our sample as a function of orbital period and planet radius. We also construct an additional sensitivity map where the orbital period is replaced by stellar insolation (the amount of stellar flux at the location of the planet's orbit) relative to Earth's value. We achieve this by converting the range of orbital periods on each individual sensitivity map into stellar insolation values using the appropriate stellar luminosity and mass for each star. Both of these sensitivity maps are provided in Figure \ref{fig:surveysens}.

We recover nearly every injected planet with a radius above 1.5 \rearth~regardless of the tested orbital period or insolation. However, average sensitivity starts to drop rapidly for planet sizes smaller than 1 Earth. Over the range of tested orbital periods, the constant sensitivity contour lines are nearly vertical, indicating that the average sensitivity is primarily a function of planet size. Note that the detection sensitivity for every star was evaluated for a fixed orbital period range (0.1-7 days). As the conversion between orbital period and insolation depends on the stellar mass and luminosity, the range of tested insolation values is different for every star, spanning values from 0.5-6.1 \searth~at $P = 7$ days (depending on the star) up to 131-1771 \searth~at $P = 0.1$ days. We elected to limit our modeling to insolations between 4-200 \searth~as the studied orbital periods cover this range for the vast majority of our targets. For 7\% of our stars, an insolation of 4 \searth~corresponds to an orbital period longer than 7 days, and this fraction starts to increase rapidly for lower insolation values. The range of insolations corresponding to the tested orbital periods (0.1-7 days) includes 200 \searth~for all but one of our targets. However, insolations above 200 \searth~correspond to orbital periods below 0.1 days. Such planets would be relatively easy to detect and are likely exceedingly rare due to their orbiting extremely close to the host stars. Therefore, even though we explicitly model an insolation range of 4-200 \searth, we effectively cover all insolations above 4 \searth.

\subsection{Parametrizing the occurrence rate}\label{sec:occparam}

We test a series of plausible parametrizations for the occurrence rate $q$ in Equation \ref{eq:occrate}. Our most general model for the period-dependent component $f(P)$ of the occurrence rate employs a broken power law:
\begin{equation}
    f(P) = \frac{\partial Q}{\partial P} = C P^\alpha \left(1 - e^{-(\frac{P}{P_0})^\gamma}\right)
\end{equation}
A similar parametrization has been used in previous works \citep{Howard2012,Mulders2018,Petigura2018} with a typical break period at $P \approx 10$ days, which is consistent with the proto-planetary disk having an inner edge near 0.1 AU \citep{Mulders2015a}. However, as our study concerns orbital periods below 7 days, the inclusion of a break point in $f(P)$ may not be statistically warranted. Therefore, we also test a simpler power law $f(P) = C P^\alpha$ with a uniform prior on $\alpha$ between -3 and 3. The latter is also consistent with the observed accepted planet distribution in Figure \ref{fig:surveysens} where the planets are distributed relatively uniformly in $\ln P$ between 0.4-7 days.

The radii of the detected planets all fall between $0.91-1.31$ \rearth. We compare multiple different models for the radius-dependent component $g(r)$ of the occurrence rate. In particular, we assume $g(r)$ is described either by a constant function or a step function with possible breaks at 0.9 \rearth~and 1.4 \rearth. The true occurrence rate is unlikely to have such sharp cutoffs as a function of radius, however, so we also include a normal distribution model for $g(r) = \No(\mu, \sigma)$. This is consistent with the observed planet distribution (Figure \ref{fig:surveysens}), corresponding to a rocky terrestrial planet population without volatile envelopes. We include Gaussian prior distributions on both $\mu$ and $\sigma$, centered at the mean (1.127 \rearth) and standard deviation (0.136 \rearth) of the observed planet radius distribution. When left unconstrained, the posterior probability distribution for $\mu$ has a long tail towards the smaller radii, representing a potential hidden planet population due to limited sensitivity centered at a radius smaller than any of the observed values. This tail has a small integrated probability while noticeably degrading the model evidence (the model with the included priors was preferred with a Bayes factor of $Z_2 / Z_1 = 8.2$, a notable improvement\footnote{In model comparison, a Bayes ratio of $K = Z_1 / Z_0$ means that the observed data are $K$ times more likely to occur under the hypothesis of the proposed model compared to the null hypothesis. A commonly used interpretation suggests that Bayes ratios between 1-3 correspond to anecdotal evidence, 3-10 to moderate evidence, and over 10 to strong evidence in favor of the proposed model. \citep{Jeffreys1961,Lee2014a}}). Therefore, we assume that the underlying radius distribution peaks close to the observed planet distribution. This is supported by the notable absence of detected planets below a radius of 0.9 \rearth in this study, despite having partial detection sensitivity in that range. The width of either prior distribution for $\mu$ or $\sigma$ is set to be 20\% of the central value. In each of the models for $g(r)$, we only optimize Equation \ref{eq:occrate2} over $0.5 R_\earth \leq r_j \leq 2 R_\earth$ since we have negligible sensitivity for radii below 0.5 \rearth.

A summary of the tested models is given in Table \ref{tbl:occmodels}, including the respective cumulative occurrence rates ($Q$) within the quoted period/insolation and radius ranges, period/insolation power law parameters ($\alpha$), and model evidence values ($\ln Z$).

\begin{deluxetable}{ccccccccc}
    \tabletypesize{\footnotesize}
	\tablewidth{0pt}
	\tablecaption{Occurrence rate models\label{tbl:occmodels}}
	\tablehead{
		No. & Model for $f(P)$ or $f(S)$ & Model for $g(r)$ & $P$ & $S$ & $r$ & $Q$ & $\alpha$ & $\ln Z$\\
		 & & & (days) & (\searth) & (\rearth) & & &
	}
	\startdata
	 1 & Broken power law & Normal & 0.4-7 & \nodata & 0.5-2 & $0.65^{+0.34}_{-0.23}$ & $0.80^{+0.72}_{-1.68}$ & $-83.463 \pm 0.156$\\
	 2 & Power law & Normal & 0.4-7 & \nodata & 0.5-2 & $0.67^{+0.33}_{-0.23}$ & $1.20^{+0.50}_{-0.46}$ & $-83.580 \pm 0.157$\\
	 3 & Linear & Normal & 0.4-7 & \nodata & 0.5-2 & $0.61^{+0.24}_{-0.19}$ & $1$ & $-81.862 \pm 0.124$\\
	 4 & Linear & Constant & 0.4-7 & \nodata & 0.5-2 & $0.63^{+0.27}_{-0.21}$ & $1$ & $-86.850 \pm 0.082$\\
	 5 & Linear & Step (break at 1.4\rearth) & 0.4-7 & \nodata & 0.5-2 & $0.87^{+0.32}_{-0.25}$ & $1$ & $-84.923 \pm 0.158$\\
	 6 & Linear & Step (break at 0.9\rearth, 1.4\rearth) & 0.4-7 & \nodata & 0.5-2 & $0.83^{+0.31}_{-0.23}$ & $1$ & $-85.405 \pm 0.191$\\
	 \hline
	 7 & Linear & Normal & 0.1-0.4 & \nodata & 0.5-2 & $\leq 0.015$ & $1$ & $-5.131 \pm 0.174$\\
	 \hline
	 8 & Linear & Constant & 0.4-7 & \nodata & 1.5-2 & $\leq 0.073$ & $1$ & $-3.465 \pm 0.141$\\
	 \hline
	 9 & Broken power law & Normal & \nodata & 4-200 & 0.5-2 & $0.53^{+0.23}_{-0.17}$ & $-1.70^{+0.44}_{-0.54}$ & $-87.227 \pm 0.163$\\
	 10 & Power law & Normal & \nodata & 4-200 & 0.5-2 & $0.53^{+0.25}_{-0.17}$ & $-1.20^{+0.35}_{-0.39}$ & $-86.907 \pm 0.163$\\
	 11 & Inverse & Normal & \nodata & 4-200 & 0.5-2 & $0.49^{+0.19}_{-0.15}$ & $-1$ & $-85.234 \pm 0.132$\\
	 \hline
	 12 & Inverse & Constant & \nodata & 4-200 & 1.5-2 & $\leq 0.060$ & $-1$ & $-3.783 \pm 0.147$\\
	\enddata
	\tablecomments{For each model, we provide the tested period range $P$ or insolation range $S$, the tested radius range $r$, the cumulative occurrence rate $Q$, the power law exponent $\alpha$ (where $f$ is modeled by a power law), and the total model evidence $\ln Z$. A linear model for $f$ ($\alpha = 1$) corresponds to a constant occurrence function $dN/dP$, where $N$ is the number of planets. Horizontal lines separate different ranges for $P$, $S$, and $r$.}
\end{deluxetable}

We constrain the optimization of the majority of our models (1-6, 8) to orbital periods between 0.4 and 7 days. This is due to the absence of detected planets with periods between 0.1-0.4 days combined with a high detection sensitivity within that range. We also note that the $a/R$ ratios corresponding to an orbital period of 0.1 days are between 2.0-2.7 (depending on stellar mass), and they are between 5.1-6.9 for $P = 0.4$ days. Meanwhile, transit probability varies between 36-49\% (for $P = 0.1$ days) and 14-20\% ($P = 0.4$ days). Thus, the lack of such planets is highly significant and breaks the overall power law for $f(P)$ optimized for the rest of the period range. We instead place an upper limit on $Q$ for $P$ between 0.1-0.4 days using a separate model (7).

Models (1) and (2) both yield a value for $\alpha$ consistent with unity, corresponding to a linear function $f(P) = \partial Q / \partial P \propto P$. This is equivalent to $\frac{\partial Q}{\partial\ln P} = const.$, a constant rate in log-period. Adopting the assumption of a linear model for $f(P)$ noticeably improves the model evidence ($Z_3 / Z_2$ = 5.6); as a result, we prefer this simpler characterization. Models (3)-(6) compare a normal distribution for $g(r)$ with step functions having varying number of steps (0, 1, or 2) to each side of the observed radius distribution of planets. Model (3), corresponding to a normal distribution, is preferred; we discuss this in detail in the following section.

Model (7) characterizes the occurrence rate separately for close-in planets with orbital periods between 0.1-0.4 days. Since there are no detected planets within that range of periods, we adopt the parametrization for $f(P)$ and $g(r)$ from model (3), the best-fitting model for periods between 0.4-7 days. Thus, we make an assumption that the occurrence rate for the shortest-period planets scales similarly in terms of $P$ and $r$, but that such planets may be intrinsically more rare. For $g(r)$, we also adopt the marginalized posterior distributions for $\mu$ and $\sigma$ from model (3) as prior distributions for the same variables in model (7).

In addition, model (8) separately evaluates the occurrence rate of enveloped terrestrials with radii 1.5-2 \rearth~by considering it as a separate distribution from the Gaussian-shaped rocky terrestrial distribution. Here, $g(r) = const.$ since we have no information about the shape of that distribution due to a lack of any detected planets with $r > 1.5$ \rearth~within the studied sample. $f(P)$ is modeled with a linear function as before.

We also include models (9)-(11) that are equivalents of models (1)-(3), but $f$ is parametrized as a function of insolation $S$ within the range of $4-200$ \searth~instead of orbital period $P$. Based on model evidence, we find that a simple inverse function ($f(S) \propto 1/S$) provides the statistically most justified model for $f(S)$. Thus, model (11) is our preferred parametrization for the occurrence rate $q(S, r)$ in terms of insolation $S$ and radius $r$.

Finally, model (12) characterizes the potential number of larger planets ($r > 1.5$ \rearth) for $S$ between 4-200 \searth, similarly to the orbital period-dependent model (8). Once again, we only use this to estimate an upper limit on the abundance of such planets due to a lack of planet detections with $r > 1.5$ \rearth.

\subsection{Cumulative occurrence rates}\label{sec:occrates}

\begin{figure}
    \includegraphics[width=\textwidth]{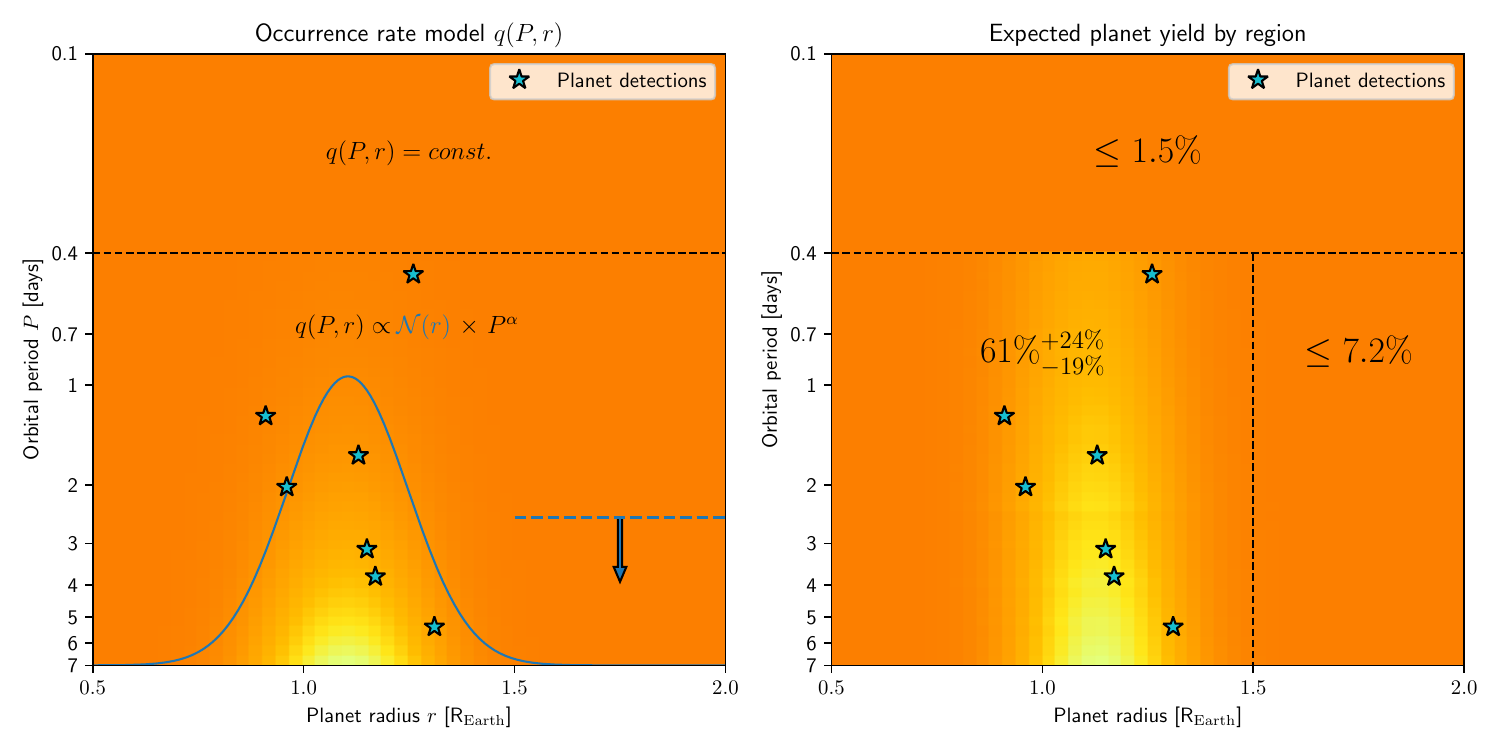}
    \caption{\textbf{Left:} The maximum a posteriori probability estimate of our underlying occurrence rate model $q(P, r)$ as a function of orbital period $P$ and planet radius $r$. Lighter colors correspond to higher probabilities. The blue solid line indicates the un-normalized radius-dependent occurrence rate $g(r)$ as a function of radius from model (3), representing the observed terrestrial planet distribution. The blue dashed line indicates the constant $g(r)$ of model (8) for $r > 1.5$ \rearth. Stars denote the locations of the seven detected planets. \textbf{Right:} The expected planet yield (likelihood of observing a planet) after combining the underlying model on the left, transit probability, and detection sensitivity as a function of orbital period $P$ and planet radius $r$. Lighter colors correspond to higher likelihoods. The quoted values are cumulative expected occurrence rates of planets per star (or $1\sigma$ upper limits) within each region bounded by the dashed lines. Stars denote the locations of the seven detected planets.}
    \label{fig:occmodel1}
\end{figure}

As demonstrated by the values in Table \ref{tbl:occmodels}, the cumulative occurrence rate $Q$ is somewhat sensitive to the chosen parametrization for $f$ and $g$, but the differences in estimated values generally remain within 1$\sigma$ levels (note that only models with identical ranges of $r$ and $S$ or $P$ can be directly compared). However, models with a higher total evidence $\ln Z$ should be preferred. For $P$ between 0.4-7 days and $r$ between 0.5-2 \rearth, model (3) has the highest evidence, with a linear parameterization for $f(P)$ (or equivalently, a constant function for $\partial Q / \partial\ln P$) and a normal distribution for $g(r)$. It yields a cumulative occurrence rate of $0.61^{+0.24}_{-0.19}$ planets per star within the quoted period and radius range. We generate marginal posterior distributions for all model parameters, finding that the Gaussian parametrization for $g(r)$ has a mean of $\mu = 1.11^{+0.05}_{-0.06}$ \rearth~and a standard deviation of $\sigma = 0.14 \pm 0.02$ \rearth. The two models (1) and (2) with a power-law-characterized $f(P)$ yield consistent occurrence rates but have a lower model evidence, supporting our decision to fix $\alpha = 1$ instead of using it as a variable. We display our best-fit occurrence rate model (3) and the expected planet yield as a function of orbital period and planet radius in Figure \ref{fig:occmodel1}.

Comparing models (3)-(6) with different parametrizations for $g(r)$, the normal distribution (model 3) is preferred with a Bayes factor of $Z_3 / Z_5 = 21$ over the best-fitting step function model (model 5), a highly significant improvement. Comparing the three step function models (4)-(6), the fifth model with a break at $r = 1.4$ \rearth~has the highest model evidence, and it is preferred over model (4) ($g(r) = const.$) with a Bayes factor of $Z_2 / Z_1 = 6.9$, also a noticeable improvement. Therefore, adding a step in $g(r)$ at $r = 1.4$ \rearth~is statistically preferred. More specifically, model (5) yields a cumulative occurrence rate of $0.72^{+0.19}_{-0.22}$ planets per star with $r$ between 0.5-1.4 Earth radii, and it places an upper limit of 0.16 planets per star with radii $r \geq 1.4R_\earth$ with 90\% confidence.

The double-step model (6) with breaks at $r = 0.9$ \rearth~and 1.4 \rearth~produces a cumulative occurrence rate of $0.13^{+0.20}_{-0.10}$ planets per star ($1\sigma$ upper limit of 0.22) for $r$ between 0.5-0.9 \rearth, and a rate of $0.59^{+0.24}_{-0.17}$ planets per star for $r$ between 0.9-1.4 \rearth. However, the difference in $\ln Z$ is negligible between models (5) and (6), suggesting that based on the current data, we cannot meaningfully reject either model for the other.

Using model (7), we place an upper limit on the occurrence rate of terrestrial planets with orbital periods between 0.1-0.4 days. We find that $Q \leq 0.015$ planets per star with 68.2\% (1$\sigma$) confidence, and $Q \leq 0.027$ planets per star with 90\% confidence. From model (8), we place a 1$\sigma$ upper limit of $Q \leq 0.073$ on the cumulative occurrence rate of enveloped terrestrials with radii 1.5-2 \rearth~and orbital periods between 0.4-7 days. However, since the average detection sensitivity approaches unity for planets with $r > 2$ \rearth~(e.g. Figure \ref{fig:surveysens}), the latter upper limit can be interpreted to apply to all sub-Neptunes with $r > 1.5$\rearth.

\begin{figure}
    \includegraphics[width=\textwidth]{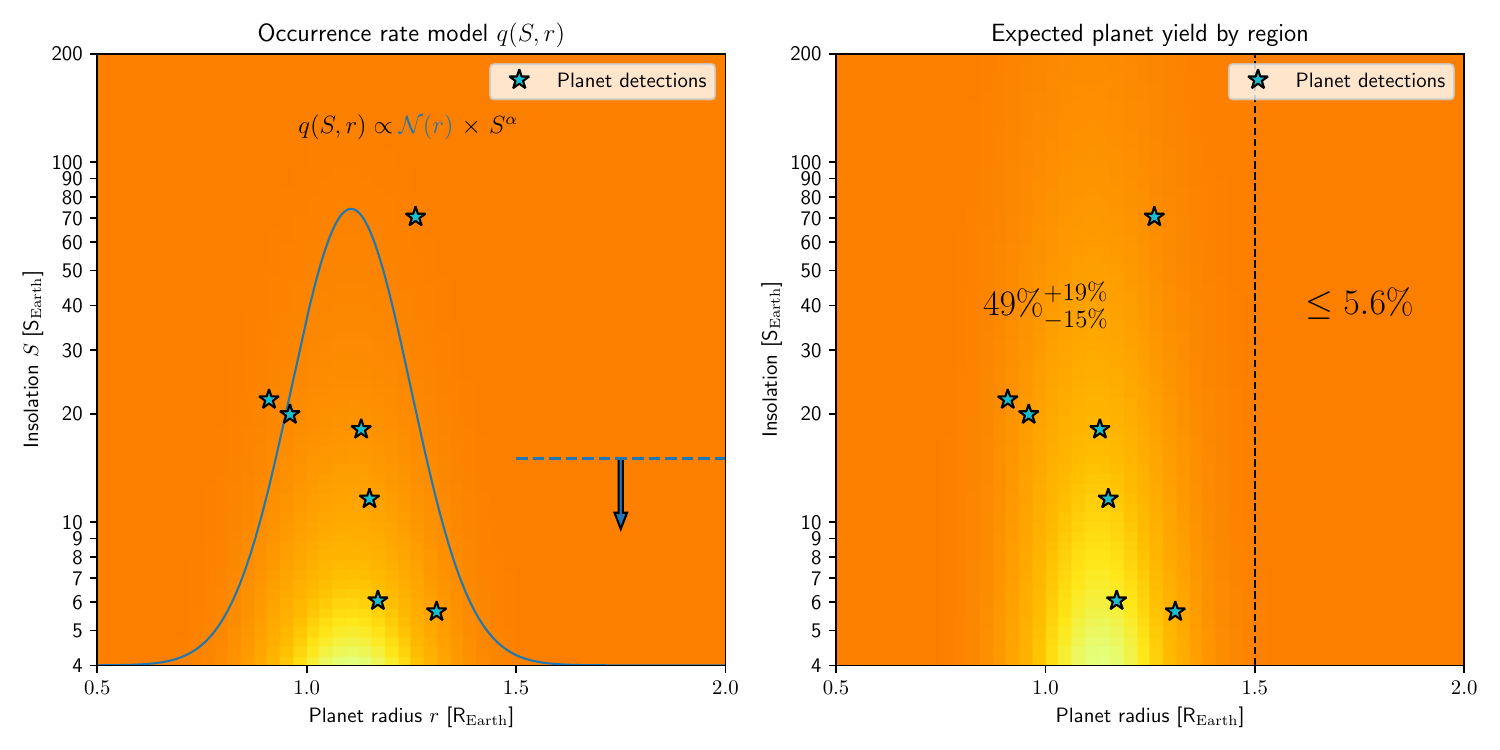}
    \caption{\textbf{Left:} The maximum a posteriori probability estimate of our underlying occurrence rate model $q(S, r)$ as a function of orbital period $P$ and planet insolation $S$. Lighter colors correspond to higher probabilities. The blue solid line indicates the un-normalized radius-dependent occurrence rate $g(r)$ as a function of radius from model (11), representing the observed terrestrial planet distribution. The blue dashed line indicates the constant $g(r)$ of model (12) for $r > 1.5$ \rearth. Stars denote the locations of the seven detected planets. \textbf{Right:} The expected planet yield (likelihood of observing a planet) after combining the underlying model on the left, transit probability, and detection sensitivity as a function of insolation $S$ and planet radius $r$. Lighter colors correspond to higher likelihoods. The quoted values are cumulative expected occurrence rates of planets per star (or $1\sigma$ upper limits) within each region bounded by the dashed lines. Stars denote the locations of the seven detected planets.}
    \label{fig:occmodel2}
\end{figure}

The preferred insolation-parametrized model (11) yields a cumulative occurrence rate of $0.49^{+0.19}_{-0.15}$ planets per star with $S$ between 4-200 \searth. As explained in Section \ref{sec:surveysens}, this tested insolation range is not exactly equivalent to the orbital period range of 0.4-7 days, which explains the insolation-parametrized models yielding slightly lower occurrence rates compared to the period-parametrized models (in addition, parametrizations for $f(P)$ and $f(S)$ are not identical). Finally, model (12) yields an estimated occurrence rate of $Q \leq 0.060$ enveloped terrestrials ($r > 1.5$ \rearth) per star, consistent with the upper limit from the period-parametrized model (8). We display the best-fit insolation-parametrized occurrence rate model (11) and the expected planet yield as a function of insolation and planet radius in Figure \ref{fig:occmodel2}.

For models (3), (7), (8), (11), and (12), we also estimate cumulative occurrence rates for various specific ranges of planet radii and orbital periods or insolations. These values are given in Table \ref{tbl:occrates}. In particular, we provide occurrence rate values integrated over the more standard radius ranges of 0.5-1 \rearth, 1-1.5 \rearth, and 1.5-2 \rearth. If the specified range has no planet detections, we provide a $1\sigma$ upper limit (68.2nd percentile) instead.

\begin{deluxetable}{cccccc}
    \tabletypesize{\footnotesize}
	\tablewidth{0pt}
	\tablecaption{Cumulative number of planets per star\label{tbl:occrates}}
	\tablehead{
		Model & \multirow{2}{*}{Par.\tablenotemark{a}} & \multirow{2}{*}{Range} & \multicolumn{3}{c}{Occ. rate by planet radius}\\
		\cline{4-6}
		No. & & & 0.5-1\rearth & 1-1.5\rearth & 1.5-2\rearth
	}
	\startdata
	 \multirow{4}{*}{3} & \multirow{4}{*}{$P$} & 0.4-1.0 days & $\leq 0.017$ & $0.041^{+0.018}_{-0.013}$ & $\leq 0.001$\\
	  & & 1.0-2.7 days & $0.035^{+0.032}_{-0.019}$ & $0.117^{+0.051}_{-0.037}$ & $\leq 0.001$\\
	  & & 2.7-7.0 days & $\leq 0.123$ & $0.295^{+0.128}_{-0.094}$ & $\leq 0.002$\\
	 \cline{3-6}
	  & & 0.5-7.0 days & $0.134^{+0.123}_{-0.072}$ & $0.446^{+0.194}_{-0.143}$ & $\leq 0.003$\\
	 \hline
	 7 & $P$ & 0.1-0.4 days & $\leq 0.003$ & $\leq 0.011$ & $\leq 0.001$\\
	 \hline
	 \multirow{2}{*}{8} & \multirow{2}{*}{$P$} & 0.4-7.0 days & \nodata & \nodata & $\leq 0.073$\\
	 \cline{3-6}
	  & & 0.5-7.0 days & \nodata & \nodata & $\leq 0.072$\\
	 \hline
	 \multirow{4}{*}{11} & \multirow{4}{*}{$S$} & 4-10 \searth & $\leq 0.090$ & $0.226^{+0.099}_{-0.072}$ & $\leq 0.002$\\
	  & & 10-50 \searth & $0.035^{+0.032}_{-0.018}$ & $0.121^{+0.053}_{-0.038}$ & $\leq 0.001$\\
	  & & 50-200 \searth & $\leq 0.009$ & $0.023^{+0.010}_{-0.007}$ & $\leq 0.001$\\
	 \cline{3-6}
	  & & 4-200 \searth & $0.107^{+0.098}_{-0.055}$ & $0.370^{+0.162}_{-0.118}$ & $\leq 0.003$\\
	 \hline
	 12 & $S$ & 4-200 \searth & \nodata & \nodata & $\leq 0.060$\\
	\enddata
	\tablenotetext{a}{$P$: orbital period, $S$: insolation.}
	\tablecomments{Model numbers correspond to models in Table \ref{tbl:occmodels}. For bins with no detected planets, we provide a $1\sigma$ (68.2-nd percentile) upper limit instead. Otherwise, the values are quoted with $1\sigma$ uncertainties.}
\end{deluxetable}

\subsection{Comparing occurrence rates with more massive M dwarfs}\label{sec:occcompare}

Table \ref{tbl:occcompare} lists our estimated cumulative occurrence rates for orbital periods of 0.5-7 days as a function of planet radius from this work and contrasts them with results from \citet[][hereon DC15]{Dressing2015}. The results of DC15 were calculated mainly based on larger M dwarfs from the full Kepler dataset. In particular, the stellar sample of DC15 had a median radius of 0.47 \rsun~(corresponding roughly to a mass of 0.50 \msun), whereas our sample of mid-to-late M dwarfs has a median radius of 0.20 \rsun~(0.17 \msun). Thus, the masses of the median star in the two studies differ nearly by a factor of three. To obtain occurrence rates for specific period and size bins, we adopt the rates published in Figure 16 of DC15 and integrate them over the required period/radius range, assuming a constant rate in radius and log-period within each bin. The quoted uncertainties are Poisson errors based on the search completeness in each bin and the sample size of DC15 (2543 stars).

With the highest model evidence, model (3) with $g(r) \sim \mathcal{N}(\mu, \sigma)$ provides the best description of the terrestrial planet distribution within the radius range of 0.5-1.5 \rearth. However, the tails of the normal distribution yield extremely small occurrence rates for planets larger than 1.5 \rearth, based on the complete absence of a detected enveloped terrestrial population in our stellar sample. In reality, we know that this population still exists: one of the stars in the original \citet{Winters2021} sample (GJ 1214) is known to possess a close-in transiting mini-Neptune \citep[$P = 1.58$ days, $r = 2.74 \pm 0.05 R_\earth$;][]{Charbonneau2009,Cloutier2021}, but it was excluded from our analysis as it has not yet been observed by TESS. Still, including GJ 1214 b would yield an occurrence rate of 0.034 enveloped terrestrials per star (assuming no other sub-Neptunes within the potentially analyzable sample of 473 stars), consistent with the upper limit of $Q \leq 0.072$ that we derive for $r > 2$ \rearth. We also note that the vast majority of our sample has been monitored by the MEarth survey \citep{Nutzman2008,Irwin2015} and has a substantial amount of ground-based photometry. Any planet that has not been detected so far is more likely to be smaller than 1.5 \rearth~based on the decreased detection sensitivity for such planets; thus, the highly lopsided occurrence rate in favor of planets with radii below 1.5 \rearth~is likely to remain.

For planets larger than 1.5 \rearth, we adopt cumulative occurrence rates from model (8) to allow for a separate but undetected population of enveloped terrestrials (with $g(r) = const.$). As detection sensitivity at any fixed orbital period increases with planet radius, and our survey sensitivity approaches 100\% for planets of radius 2 \rearth, we can also estimate the occurrence rates of any planet larger than 2 \rearth~by adopting the detection sensitivity at 2 \rearth~for any planet with $r > 2$ \rearth. As both radius ranges have no detected planets, we find a negligible difference between the occurrence rates at 1.5-2 \rearth~and above 2 \rearth, with the uncertainties driven solely by transit probability and the limited sample size in both cases.

Overall, we find a cumulative occurrence rate of $0.61^{+0.24}_{-0.19}$ planets per star (pps) between periods of 0.5-7 days and planet radii of 0.5-4 \rearth, slightly above but consistent with the $0.47^{+0.07}_{-0.06}$ pps from DC15. However, there are notable differences in the planet distribution as a function of radius. For the smallest planets (radii between 0.5-1 \rearth), we calculate an incidence of $0.134^{+0.123}_{-0.072}$ pps, consistent with the value of $0.132^{+0.047}_{-0.041}$ pps by DC15. For the radius range of 1.0-1.5 \rearth, we find a much higher occurrence rate of $0.446^{+0.194}_{-0.143}$ pps compared to the $0.157^{+0.036}_{-0.032}$ pps by DC15. In contrast, we find $Q \leq 0.073$ pps for $1.5 < r/R_\earth < 2$ and $Q \leq 0.072$ pps for $r > 2$ \rearth, slightly lower but mostly consistent with the respective values of $Q = 0.074^{+0.022}_{-0.020}$ pps and $Q = 0.103^{+0.025}_{-0.024}$ pps by DC15.

We analogously derive and compare occurrence rates in bins of insolation versus planet radius based on the model with the highest evidence (model 11). These results are provided in Table \ref{tbl:occcompare2}. For planets larger than 0.5 \rearth~and insolation between 4-200 \searth, we find an occurrence rate of $0.49^{+0.19}_{-0.15}$ pps, substantially smaller than the equivalent $0.99^{+0.16}_{-0.15}$ pps from DC15. This much larger discrepancy compared to the period-integrated values in Table \ref{tbl:occcompare} is supported by the fact that for a given insolation, planets orbiting more massive stars will have a longer orbital period, and the occurrence rate as a function of $\ln P$ has a positive slope in DC15 for periods up to about 25 days. Breaking the occurrence rate up as a function of planet radius, we find a lower occurrence rate of $0.107^{+0.098}_{-0.055}$ pps compared to $0.233^{+0.122}_{-0.097}$ pps by DC15 for planet radii between 0.5-1 \rearth. For radii between 1-1.5 \rearth, our occurrence rate of $0.370^{+0.162}_{-0.118}$ pps is somewhat higher than the $0.198^{+0.066}_{-0.057}$ pps in the more massive M dwarf sample of DC15. However, the most significant difference between the two samples occurs for larger planets: we estimate $Q \leq 0.060$ pps and $Q \leq 0.056$ pps for $1.5 < r/R_\earth < 2$ and $r > 2$ \rearth, respectively, while the equivalent values by DC15 are $Q = 0.198^{+0.066}_{-0.057}$ pps and $Q = 0.283^{+0.074}_{-0.067}$ pps. Finally, we note that the contribution to the occurrence rate from planets with insolations greater than 200 \searth~is negligible in both samples, and we can adopt the values for $S$ between 4-200 \searth~to represent the overall planet population with $S \geq 4$ \searth.

\begin{deluxetable}{cccc}
    \tabletypesize{\footnotesize}
	\tablewidth{0pt}
	\tablecaption{Comparison of cumulative occurrence rates for various orbital periods\label{tbl:occcompare}}
	\tablehead{
		Period & Radius & \multicolumn{2}{c}{Occ. rate by source}\\
		\cline{3-4}
		(days) & (\rearth) & DC15\tablenotemark{a} & This work\\
		\multicolumn{2}{r}{Median stellar mass:} & 0.50 \msun & 0.17 \msun
	}
	\startdata
	 $0.5-1$ & \multirow{3}{*}{$0.5-4.0$} & $0.013^{+0.005}_{-0.005}$ & $0.047^{+0.019}_{-0.015}$\\
	 $1-2.7$ & & $0.106^{+0.024}_{-0.024}$ & $0.159^{+0.064}_{-0.050}$\\
	 $2.7-7$ & & $0.349^{+0.054}_{-0.051}$ & $0.402^{+0.162}_{-0.127}$\\
	 \hline
	 \multirow{5}{*}{$0.5-7$} & $0.5-1.0$ & $0.132^{+0.047}_{-0.041}$ & $0.134^{+0.123}_{-0.072}$\\
	  & $1.0-1.5$ & $0.157^{+0.035}_{-0.032}$ & $0.446^{+0.194}_{-0.143}$\\
	  & $1.5-2.0$ & $0.074^{+0.022}_{-0.020}$ & $\leq 0.073$\tablenotemark{b}\\
	  & $>2.0$ & $0.103^{+0.025}_{-0.024}$ & $\leq 0.072$\tablenotemark{c}\\
	  \cline{2-4}
	  & $0.5-4.0$ & $0.47^{+0.07}_{-0.06}$ & $0.61^{+0.24}_{-0.19}$\\
	\enddata
	\tablenotetext{a}{From \citet{Dressing2015}.}
    \tablenotetext{b}{This radius range had no planet detections. We provide a 1$\sigma$ (68.2\%) upper limit instead.}
	\tablenotetext{c}{This radius range was not directly simulated. We provide a 1$\sigma$ (68.2\%) upper limit instead, based on our detection sensitivity at 2\rearth.}
\end{deluxetable}
\begin{deluxetable}{cccc}
    \tabletypesize{\footnotesize}
	\tablewidth{0pt}
	\tablecaption{Comparison of cumulative occurrence rates for various planet insolations\label{tbl:occcompare2}}
	\tablehead{
		Insolation & Radius & \multicolumn{2}{c}{Occ. rate by source}\\
		\cline{3-4}
		(\searth) & (\rearth) & DC15\tablenotemark{a} & This work\\
		\multicolumn{2}{r}{Median stellar mass:} & 0.50 \msun & 0.17 \msun
	}
	\startdata
	 $4-10$ & \multirow{3}{*}{$0.5-4.0$} & $0.436^{+0.140}_{-0.113}$ & $0.303^{+0.120}_{-0.095}$\\
	 $10-50$ & & $0.438^{+0.096}_{-0.083}$ & $0.162^{+0.064}_{-0.051}$\\
	 $50-200$ & & $0.084^{+0.024}_{-0.022}$ & $0.030^{+0.012}_{-0.010}$\\
	 \hline
	 \multirow{5}{*}{$4-200$} & $0.5-1.0$ & $0.233^{+0.122}_{-0.097}$ & $0.107^{+0.098}_{-0.055}$\\
	  & $1.0-1.5$ & $0.243^{+0.069}_{-0.068}$ & $0.370^{+0.162}_{-0.118}$\\
	  & $1.5-2.0$ & $0.198^{+0.066}_{-0.057}$ & $\leq 0.060$\tablenotemark{b}\\
	  & $>2.0$ & $0.283^{+0.074}_{-0.067}$ & $\leq 0.056$\tablenotemark{c}\\
	  \cline{2-4}
	  & $0.5-4.0$ & $0.99^{+0.16}_{-0.15}$ & $0.49^{+0.19}_{-0.15}$\\
	\enddata
	\tablenotetext{a}{From \citet{Dressing2015}.}
	\tablenotetext{b}{This radius range had no planet detections. We provide a 1$\sigma$ (68.2\%) upper limit instead.}
	\tablenotetext{c}{This radius range was not directly simulated. We provide a 1$\sigma$ (68.2\%) upper limit instead, based on our detection sensitivity at 2\rearth.}
\end{deluxetable}

\subsection{Occurrence rates by planet type}\label{sec:occbytype}

All of the detected planets in this study have radii below 1.31 \rearth~and bulk densities that are either confirmed or highly likely to be consistent with a predominantly rocky composition without a notable volatile envelope. Thus, we take this planet population and the underlying incidence rate model for $q(P,r)$ (e.g. Figure \ref{fig:occmodel1}) to be representative of the rocky planet population. We assign planet radii $r > 1.5$ \rearth~to represent the population of sub-Neptunes (including enveloped terrestrials and water worlds) and model the underlying occurrence rate to be constant with radius ($\delta q / \delta r = 0$) since this population contains no planet detections in this study. While the exact location of the radius valley is unknown for mid-to-late M dwarfs, the choice of $r = 1.5$ \rearth~separating the super-Earth and sub-Neptune populations is consistent with the estimated location of the radius valley at $r = 1.60^{+0.34}_{-0.35}$ \rearth~for early M dwarfs \citep[median mass 0.5 \msun;][]{Cloutier2020}. It is also a relatively safe choice since the modeled rocky planet occurrence rate model $q(P, r)$ drops close to zero at $r = 1.5$ \rearth~and detection sensitivity remains relatively constant for $r > 1.5$ \rearth; thus, moving the location of the radius valley to any radius between 1.5-2 \rearth~will not have a significant effect on the calculated occurrence rates for either population.

We provide estimated occurrence rates for rocky planets and sub-Neptunes in Table \ref{tbl:occbytype}. Analogously to Section \ref{sec:occcompare}, we compare them to the values published by DC15 for early-to-mid M dwarfs. For orbital periods between 0.5-7 days, we find an elevated occurrence rate of $0.61^{+0.24}_{-0.19}$ rocky planets per M dwarf compared to the value of $0.29 \pm 0.05$ pps by DC 15. However, the abundance of sub-Neptunes is substantially suppressed around mid-to-late M dwarfs: we set a 1$\sigma$ upper limit of 0.07 pps versus the value of $0.18 \pm 0.04$ by DC15. When taking into account planet insolation, we find $0.49^{+0.19}_{-0.15}$ rocky planets per M dwarf with $S \geq 4$ \searth, consistent with the value of $0.48^{+0.14}_{-0.13}$ pps by DC15. The discrepancy in the abundance of sub-Neptunes is even more significant: we obtain a 1$\sigma$ upper limit of 0.06 pps compared to the much higher value of $0.49 \pm 0.10$ pps for early-to-mid M dwarfs by DC15.

\begin{deluxetable}{ccccc}
    \tabletypesize{\footnotesize}
	\tablewidth{0pt}
	\tablecaption{Comparison of cumulative occurrence rates by planet type\label{tbl:occbytype}}
	\tablehead{
		\multirow{2}{*}{Parameter} & \multirow{2}{*}{Range} & \multirow{2}{*}{Planet type\tablenotemark{a}} & \multicolumn{2}{c}{Occ. rate by source}\\
		\cline{4-5}
		 & & & DC15\tablenotemark{b} & This work\\
		\multicolumn{3}{r}{Median stellar mass:} & 0.50 \msun & 0.17 \msun
	}
	\startdata
	 \multirow{3}{*}{Orbital period} & \multirow{3}{*}{$0.5-7$ days} & Rocky planets & $0.29 \pm 0.05$ & $0.61^{+0.24}_{-0.19}$\\
	  & & Sub-Neptunes & $0.18 \pm 0.04$ & $\leq 0.07$\\
        \cline{3-5}
        & & Est. ratio\tablenotemark{c} & 1.6 : 1 & 14 : 1\\
      \hline
      \multirow{3}{*}{Insolation} & \multirow{3}{*}{$\geq 4$ \searth} & Rocky planets & $0.48^{+0.14}_{-0.13}$ & $0.49^{+0.19}_{-0.15}$\\
	  & & Sub-Neptunes & $0.49 \pm 0.10$ & $\leq 0.06$\\
        \cline{3-5}
        & & Est. ratio\tablenotemark{c} & 1 : 1 & 13 : 1\\
	\enddata
    \tablenotetext{a}{Planets with radii $r < 1.5$ \rearth~are assumed to be rocky while planets with $r > 1.5$ \rearth~represent sub-Neptunes. See Section \ref{sec:occbytype} for explanation.}
	\tablenotetext{b}{From \citet{Dressing2015}.}
    \tablenotetext{c}{Estimated relative prominence of rocky planets compared to sub-Neptunes.}
\end{deluxetable}

\section{Discussion}

\subsection{Notable trends}

Multiple notable trends emerge regarding the population of close-in planets orbiting mid-to-late M dwarfs. While our best model predicts an occurrence rate of $0.61^{+0.24}_{-0.19}$ planets with periods less than 7 days, the observed planet population consists solely of planets smaller than 1.31 \rearth, all of which have bulk densities that are consistent with an Earth-like rocky composition. By contrast, we calculate a $1\sigma$ upper limit of 0.07 planets larger than 1.5 \rearth per star, with the 50th percentile (mean) of the posterior probability distribution at 0.043 planets per star (pps). Therefore, we estimate that rocky planets outnumber enveloped terrestrials (and water worlds) by a factor of 14 to 1. If we adopt an enveloped terrestrial occurrence rate of 0.034 pps based on GJ 1214 b, the only known such planet in the sample of stars within 15 pc with masses between 0.1-0.3 \msun, we obtain an ever larger ratio of 18 to 1. Repeating the calculation for planet insolations $S \geq 4$ \searth~yields a similar ratio of 13 to 1.

This lopsided ratio in favor of terrestrial planets is further reinforced by the near-complete sensitivity for sub-Neptunes (planets with $r>1.5R_\earth$) in this study, as demonstrated by Figure \ref{fig:surveysens}. While we were able to analyze 363 out of 512 stars in this volume-complete, all-sky data set, there are 110 additional mid-to-late M dwarfs that have not yet been observed by TESS, in addition to the 39 M dwarfs that are close companions to more massive stars and could not be deblended. Of the remaining unobserved stars, many have been extensively monitored by the ground-based MEarth survey which was responsible for the discovery of GJ 1214 b \citep{Charbonneau2009}. Despite detection sensitivity favoring the discovery of larger planets, all 5 original TESS planet discoveries described in this study (TOI 540 b, LHS 475 b, LHS 3844 b, LTT 1445A b and c) have radii below 1.5\rearth~and are likely to be rocky terrestrials. Subsequently, any remaining undiscovered planets are more probable to be terrestrial, and rocky terrestrials will likely continue to vastly outnumber sub-Neptunes close to M dwarfs with masses between 0.1-0.3 \mearth. We also note that the well-known TRAPPIST-1 multi-planet system is excluded from our sample on account of the stellar mass being below our threshold ($M = 0.08 M_\sun$), but it contains four known planets (b, c, d, and e) with orbital periods below 7 days, all established to be terrestrial with estimated radii between 0.79-1.12 \rearth~\citep{Agol2021}.

Our results fit into a larger trend of increasing relative prominence of terrestrials compared to non-rocky planets with decreasing stellar mass. For insolations above 10 \searth, sub-Neptunes outnumber terrestrial planets overall around Sun-like stars based on data from the Kepler mission \citep{Fulton2017}. However, binning the planet population as a function of stellar mass reveals a dwindling occurrence of sub-Neptunes around low-mass stars; specifically, \citet{Cloutier2020} estimated a ratio of $8.5 \pm 4.6$ of rocky to non-rocky planets for stellar masses between 0.08 \msun~and 0.42 \msun. The bimodal radius distribution of planets identified by \citet{Fulton2017} all but disappears for stars with $M < 0.42$ \msun, instead appearing nearly unimodal \citep{Cloutier2020}. The apparent unimodality of the radius distribution is present in this work as well and further justifies our choice of a single Gaussian distribution to model the occurrence rate as a function of radius (Figures \ref{fig:occmodel1} and \ref{fig:occmodel2}). We also find an average terrestrial planet size of 1.1 \rearth~around mid-to-late M dwarfs, below the average terrestrial planet radius of 1.3 \rearth~around Sun-like stars \citep{Fulton2018}. This is not surprising as the location of the peak of the rocky planet distribution as a function of radius shifts towards lower radius values with decreasing stellar mass based on studies of the Kepler sample for stellar masses above 0.3 \msun~\citep{Fulton2018,Wu2019,Cloutier2020}.

The median surface insolation of the planets in our sample is 18 \searth, and 6 out of the 7 insolation values fall within the range of 5-22 \searth. Similarly, the expected median insolation of the terrestrial planet population from model (11) is 20 \searth. The relative absence of sub-Neptunes orbiting the mid-to-late M dwarfs in this study, in contrast with heavier stars, is not a consequence of the relatively high ($S \geq 4$ \searth) incident stellar flux values in our studied parameter space. The sub-Neptune populations in similar studies of hotter stars have typical insolations equal to or greater than the planets studied here. Indeed, the sub-Neptunes detected in the Kepler sample have average insolations of 20-60 \searth~for Sun-like stars \citep{Fulton2018} and 5-30 \searth~for earlier M dwarfs \citep{Cloutier2020}. Similarly, GJ 1214 b \citep[$r = 2.74 \pm 0.05$ \rearth][]{Cloutier2021} has an estimated insolation value of 16 \searth. Thus, the heavily suppressed abundance of sub-Neptunes orbiting mid-to-late M dwarfs is an intrinsic feature of the planet population, at least for orbital periods below 7 days.

We also do not detect a population of planets substantially smaller than Earth, despite having partial sensitivity to such planets. This suggests that if a population of sub-Earths exists around mid-to-late M dwarfs, such planets are likely less abundant than Earth-sized and larger planets, or that this distribution occurs at smaller planet sizes where we have limited sensitivity. Instead, we do find tentative evidence of a downturn in planet occurrence for radii between 0.5-0.9 \rearth~compared to planets with radii between 0.9-1.3 \rearth. Specifically, we calculate a $1\sigma$ upper limit of 0.21 planets per star with sizes between 0.5-0.9 \rearth~and periods below 7 days, much lower than our estimated occurrence rate of 0.56 pps obtained by integrating $q(P,r)$ over the radius range of 0.9-1.4 \rearth. We also note that all 7 planets within this sample were detected at very high SNR' values (Table \ref{tbl:planets}), much higher than our detection limit SNR' of 6.39 or 6.55. This supports the interpretation that the lack of planets with $r < 0.9$ \rearth~is not purely a result of decreased detection sensitivity; if we were limited by sensitivity, we would expect more TCEs with low SNR' values.

The cumulative occurrence rate of $0.61^{+0.24}_{-0.19}$ pps with orbital periods 0.5-7 days and radii above 0.5\rearth~is consistent with a similar estimate of $0.47^{+0.07}_{-0.06}$ pps for earlier M dwarfs by \citet{Dressing2015}. It is also similar to the $1.06^{+0.35}_{-0.28}$ pps estimated by \citet{Sabotta2021} based on RV data from CARMENES for stellar masses below 0.34 \msun, planet minimum masses between 1-10 \mearth, and orbital periods between 1-10 days; however, the detection completeness for masses below 2\mearth~was very low within that study. We also note that accumulating additional 2-minute cadence data from TESS will likely show diminishing returns for discovering additional planets, with the vast majority of planet discoveries expected to come from the first 4 years of the TESS mission \citep{Kunimoto2022}. However, additional discoveries may come from full frame image searches of mid-to-late M dwarfs beyond 15 pc that are not included in this volume-complete sample.

We do not detect an increase in the occurrence rate with decreasing stellar mass within the M dwarf regime: the median stellar mass of planet-hosting stars in our sample is 0.18 \msun, indistinguishable from the median mass of the entire sample (0.17 \msun). This is in contrast with the Kepler study by \citet{HardegreeUllman2019} that found a sharply increasing occurrence rate towards later spectral types (between M3V, M4V, and M5V spectral types). In fact, three of the six planet hosting stars in this study are classified as M3.0V or M3.5V, the earliest of the three spectral types, despite stars of later spectral types being more numerous. However, the number of planets in both samples is likely too small to draw any definitive conclusions. When compared to the early-to-mid M dwarf sample of \citet{Dressing2015}, we instead find that the cumulative occurrence rate of all planets within our mid-to-late M dwarf sample is consistent with their work within the same period range (0.5-7 days), and lower by a factor of two when considering an equivalent range of insolations (\textgreater 4 \searth), as demonstrated in Tables \ref{tbl:occcompare} and \ref{tbl:occcompare2}. This is consistent with the findings of \citet{Brady2022} concluding that the planet occurrence rate likely does not increase and may decrease for the latest M dwarfs. It is also consistent with the recent RV study from \citet{Sabotta2021} that did not find evidence in favor of planets orbiting stars with masses below 0.34\msun~being more common than planets orbiting stars with masses above 0.34\msun, for orbital periods between 1-100 days, although the survey had a low detection sensitivity to most planets below several Earth masses.

\subsection{Broader context in terms of planet formation and evolution}

Accurately characterizing the abundance and properties of planets by host star mass yields interesting clues about various proposed planet formation and evolution scenarios. Protoplanetary disk mass correlates positively with protostar mass \citep{Pascucci2016}, and therefore one might expect planets to be more abundant around massive stars. However, contrary to expectations, M dwarfs have been shown to be more efficient at making planets than heavier stars \citep{Dressing2015}. One potential explanation invokes planet formation by pebble drift and accretion. Cold gas giants can limit the flow of pebbles into the inner protoplanetary disk, thereby inhibiting the formation of super-Earths \citep{Lambrechts2019,Mulders2021}. The abundance of gas giants decreases noticeably for M dwarfs, and the relatively low total disk mass may begin to inhibit super-Earth formation around late M dwarfs. Pebble accretion is a relatively inefficient way of forming planets \citep[i.e. most of the solid material ends up falling into the star;][]{Lin2018}; thus, a downturn in planet abundance rates for late M dwarfs would support pebble accretion as an important planet formation mechanism. Meanwhile, a higher planet formation efficiency would favor planetesimal accretion over pebble accretion \citep{Lu2020}. The results of this study are consistent with the hypothesis that late M dwarfs do not form more planets than early M dwarfs, and the cumulative planet occurrence rate may even be lower for late M dwarfs when comparing planets of similar insolation.

The heavily suppressed abundance of sub-Neptunes around mid-to-late M dwarfs analyzed in this study could be a sign of increased efficiency of atmospheric erosion due to photoevaporation close to these stars. Mid-to-late M dwarfs stay in their XUV active evolutionary phase longer \citep{Schneider2018} and emit a larger fraction of their total energy in XUV wavelengths compared to heavier stars \citep[e.g.][]{Wright2018}. However, while photoevaporation and core-powered mass loss have been proposed to be the main mechanisms sculpting the planet radius valley in Sun-like stars, the small planet population around M dwarfs is more consistent with being driven by formation in a gas-depleted environment \citep{Cloutier2020}. Thus, it is also possible that many of the terrestrial planets described in this work are primordially rocky as opposed to being stripped cores of sub-Neptunes. Distinguishing between the two hypotheses requires additional study within a larger sample size in order to determine how the location of the radius valley changes with planet insolation \citep{Cloutier2020} or orbital period \citep{Lopez2018}. The population of sub-Neptunes also includes planets with mean densities consistent with a 1:1 proportion of rock and water ice by mass. Such planets are typically referred to as "water worlds" and have lower mean densities compared to purely rocky planets \citep{Luque2022}. Simulations \citep[e.g.][]{Burn2021} predict that water worlds should be more common around low-mass stars due to faster type I migration of such planets. For the subset of planets described within this work that have measured masses (GJ 1132 b, LHS 1140 c, LTT 1445A b and c), all are consistent with a purely rocky composition, and the rest would be too small to exceed the minimum mass requirement for water worlds \citep{Burn2021}. Thus, our results suggest that mid-to-late M dwarfs either form fewer water worlds beyond the snow line compared to more massive stars or that type I migration does not efficiently move such planets to orbital periods below 7 days.

Regarding the lack of planets smaller than 0.9\rearth, we find a contrast with recent studies of GK dwarfs that have modeled a separate population of sub-Earths ($r < 1$ \rearth) which appears to increase in prominence with decreasing planet radius \citep{Hsu2019,Neil2020,Qian2021}. These studies have been motivated by a hypothesis that the rocky planet population is in fact a combination of evaporated cores of initial sub-Neptunes (super-Earths) and primordially rocky planets (sub-Earths) \citep{Neil2020,Qian2021}. We do not detect a bimodal rocky planet distribution in this study, although it is possible for another population of planets to emerge for radii much lower than 0.9\rearth~where our sensitivity decreases substantially. However, it is also plausible that typical rocky planet sizes around M dwarfs may in fact be concentrated around 1\rearth. The radius peak of the rocky planet distribution around low-mass stars appears to converge towards that value \citep{Cloutier2020}. Pebble accretion simulations predict that masses of hot super-Earths and cold gas giants are anti-correlated in systems that harbor both types of planets, providing an explanation for the lack of super-Earths in the Solar System \citep{Mulders2021}. The absence of massive gas giants in M dwarf systems would facilitate the formation of more massive close-in rocky planets. Furthermore, final core masses depend heavily on the availability of heavy elements which is higher in M dwarfs compared to GK dwarfs \citep{Mulders2015,Neil2018}. The results presented in this study support the conclusion that the relative abundance of super-Earths compared to sub-Earths may be higher in M dwarfs compared to more massive stars like the Sun.

\acknowledgments
This paper includes data collected by the TESS mission, which are publicly available from the Mikulski Archive for Space Telescopes (MAST). All the TESS data used in this paper can be found in MAST: \dataset[10.17909/jc9t-3m87]{http://dx.doi.org/10.17909/jc9t-3m87}. Funding for the TESS mission is provided by NASA’s Science Mission directorate. We acknowledge the use of public TESS Alert data from pipelines at the TESS Science Office and at the TESS Science Processing Operations Center. We would like to thank Ryan Cloutier for helpful conversations. This research made use of \textit{exoplanet} \citep{ForemanMackey2019} and its dependencies \citep{exoplanet:agol19,exoplanet:astropy13,exoplanet:astropy18,ForemanMackey2017,ForemanMackey2018,exoplanet:luger18,exoplanet:pymc3,exoplanet:theano}.

\facilities{TESS}

\bibliography{main}{}
\bibliographystyle{aasjournal}



\end{document}